# Reservoir-Engineered Refrigeration of a Superconducting Cavity with Double-Quantum-Dot Spin Qubits

Daryoosh Vashaee[a,b*] and Jahanfar Abouie[c]

[a] Department of Materials Science & Engineering, North Carolina State University, Raleigh, NC 27606, USA

[b] Department of Electrical & Computer Engineering, North Carolina State University, Raleigh, NC 27606, USA

[c] Department of Physics, Institute for Advanced Studies in Basic Sciences, Zanjan 45137-66731, Iran

We present an analytically tractable theory of reservoir-engineered refrigeration of a superconducting microwave cavity and map it onto a realistic solid-state implementation based on gate-defined double-quantum-dot (DQD) spin qubits. Treating the DQD not as a spectroscopic element but as a tunable engineered reservoir, we show how gate control of populations, coherences, linewidths, and detuning defines an effective photon birth–death process with predictable detailed balance. This framework yields closed-form expressions for the cavity steady state, identifies cooling bounds and detuning-dependent refrigeration valleys, and clarifies when refrigeration can drive the cavity below both the bath temperature and the DQD setpoint. By distinguishing refreshed (collision-like) and persistent reservoir regimes, we show how memory effects, saturation, and dark-state formation constrain cooling in realistic devices, while collective bright-mode coupling in a two-dot configuration can enhance refrigeration subject to mismatch and dephasing, as confirmed by numerical Lindblad simulations demonstrating targeted millikelvin cavity cooling relevant for cryogenic circuit-QED architectures.

## 1    Introduction

Scaling quantum hardware is increasingly constrained not only by qubit coherence, but by cryogenic heat budgets and the need to thermally stabilize microwave and mesoscopic subsystems. In many architectures, cooling power decreases sharply below the kelvin scale, while the number of electromagnetic modes, qubits, interconnects, and proximal control circuitry that must be accommodated continues to grow.[1,2] This creates a practical bottleneck: even when coherence is adequate, the coldest stage cannot necessarily host all elements that benefit from deep refrigeration. A complementary route is therefore to engineer localized nonequilibrium "cold spots" that selectively cool a targeted cavity mode or device degree of freedom while the surrounding environment remains at a higher temperature.

Here we develop an analytically tractable theory for such targeted refrigeration in circuit QED and map it onto a realistic solid-state implementation based on gate-defined DQDs. The key point is that we do not treat the DQD-cavity system primarily as a spectroscopy platform. Instead, we treat the DQD as an engineered, tunable reservoir. Its steady-state properties (populations, coherences, homogeneous linewidth, and spectral overlap with the cavity) are controlled via gates and dissipation. This engineered interaction defines an effective photon birth-death process with a predictable steady-state detailed balance. This "reservoir viewpoint" puts the refrigeration metrics (effective cavity temperature, cooling depth, and robustness to leakage and broadening) on the same footing as the device control knobs: DQD-cavity coupling, detuning, inter-dot exchange, dephasing, and the effective cavity energy damping rate $\kappa$.

A central technical ingredient is an effective DQD two-spin Hamiltonian, which captures experimentally relevant exchange coupling, anisotropies, and spin-orbit-enabled terms while remaining analytically manageable. From this Hamiltonian we derive closed-form expressions for the reservoir-induced upward and downward transition weights that enter the cavity's Lindblad rates, making the refrigeration problem reducible to analytic steady states for the cavity photon number and temperature. In this way, gate control and material-dependent terms enter the refrigeration performance through explicit, interpretable combinations rather than as hidden numerical parameters.

---

[*] dvashae@ncsu.edu

This framework clarifies why our results extend beyond standard treatments of DQD–cavity coupling. Whereas typical studies emphasize dispersive shifts, strong-coupling regimes, or qubit spectroscopy,[3,4] we focus squarely on the system's function as a refrigerator. Our emphasis is on (i) analytic steady states and explicit cooling conditions, (ii) the competition between the engineered reservoir and environmental loading, and (iii) identifying regimes in which refrigeration is either fundamentally bounded or can become genuinely sub-setpoint. This perspective is essential for assessing the DQD's viability as a functional thermal primitive within a larger cryogenic architecture.

Conceptually, we build a controlled bridge between two paradigms that are often treated separately. In the refreshed-reservoir (collision) picture, a Poissonian stream of reset ancillae interacts weakly and repetitively with the cavity, producing Markovian photon birth-death dynamics with transparent analytic steady states.[5,6] In the persistent-emitter regime relevant to solid-state implementations, the emitter remains coupled over many cavity exchange events and can accumulate correlations and energy, so refrigeration can be limited by memory and saturation.

We make this connection quantitative by identifying the collision-model strength with the on-resonance ($\omega_{qd} - \omega_c = 0$) cavity-DQD energy-exchange rate $\Gamma_c(0)$ (set by the cavity-DQD coupling $g$ and the homogeneous linewidth $\kappa + \gamma_\perp$), and by showing that a fast clamp/reset of the DQD populations restores the separation of timescales needed for stream-like statistics. This yields a practical device-level criterion in terms of the exchange-to-leakage ratio $\Gamma_c(0)/\kappa$: when $\Gamma_c(0) \gtrsim \kappa$ and the DQD is rapidly re-prepared (reset/clamp rates $\gg \{\Gamma_c, \kappa\}$), the persistent device reproduces the refreshed collision-reservoir behavior; when $\Gamma_c(0) \ll \kappa$ or re-preparation is slow, deviations arise from residual memory.[†]

Finally, we extend the analysis beyond the single-emitter limit by considering a two-emitter (both-dots-active) configuration in which collective bright/dark channels emerge. In this setting, intra-pair correlations renormalize the effective upward and downward transition rates presented to the cavity, providing a solid-state analogue of correlation-assisted refrigeration: the cavity can be driven below the reset/setpoint temperature near resonance, which is not possible in the one-active-dot geometry for the same baseline parameters. We quantify the robustness of this collective advantage against realistic inter-dot frequency mismatch ($\Delta_{12}$) and dephasing, including both symmetry-protected even-in-$\Delta_{12}$ behavior in the two-active-dot case and sign-dependent degradation in the one-active-dot case. The resulting design rules delineate when sub-100 mK cavity steady states are achievable even when the ambient environment is at the kelvin stage, and they identify which control knobs most efficiently trade off cooling depth, bandwidth, and tolerance to nonidealities

## 2   Scope and relation to the collision model

This section develops a persistent-emitter theory for refrigeration using gate-defined DQDs coupled to a superconducting cavity, and relates it to our previous study of collision model.[6] The collision model treats the reservoir as a Poisson stream of internally correlated two-level systems (TLS) that interact briefly with the cavity before being reset. Here we analyze the complementary, always-on regime where a DQD remains persistently coupled to both the cavity and its phonon environment. We treat two device-level scenarios: (i) single-emitter, where one dot of the DQD is the active emitter coupled to the cavity; and (ii) two-emitter, where both dots couple to the same cavity mode (a Tavis-Cummings-type configuration). The key new elements are: (a) a microscopic mapping from a realistic DQD Hamiltonian (including SOC-induced anisotropic exchange) to the effective transverse Jaynes-Cummings coupling $g \equiv g_\perp$ (Sec. 3.2) and the cavity-emitter detuning $\Delta \equiv \omega_{qd} - \omega_c$; (b) a controlled weak-coupling reduction that incorporates homogeneous broadening through the total transverse decoherence rate $\gamma_\perp \equiv \gamma_1/2 + \gamma_\phi$ (with $\gamma_1$ the longitudinal relaxation rate and $\gamma_\phi$ the pure-dephasing rate, defined below); and (c) the two-emitter extension, which exhibits bright/dark structure and correlation-assisted cooling, providing the solid-state analogue of the pair-coherence enhancement in the collision model.

---

[†] Note that $\Gamma_c(0) = 4g^2/(\kappa + \gamma_\perp)$ already contains $\kappa$ and $\gamma_\perp$; inequalities such as $\Gamma_c(0) \gtrsim \kappa$ or $\Gamma_c(0) \gtrsim \gamma_1$ are therefore shorthand for the equivalent scale-separation conditions written in terms of independent knobs: $\Gamma_c(0)/\kappa = 4g^2/[\kappa(\kappa + \gamma_\perp)] \gtrsim 1$ (engineered exchange dominates cavity loading), and $4g^2 \gtrsim \gamma_1(\kappa + \gamma_\perp)$ (exchange outpaces DQD rethermalization). We retain $\Gamma_c$ because it is the experimentally inferred on-resonance energy-exchange rate (e.g., from cavity linewidth change or DQD spectroscopy), while the validity condition for the secular/overlap reduction remains $g \ll \kappa + \gamma_\perp$. These quantities are defined later in the text.

In terms of the conceptual bridge (persistent DQD vs. stream), the collision model is Markovian by construction at the coarse-grained level: each prepared pair arrives uncorrelated with the cavity, interacts weakly for a short time $\tau$ (with $\phi \equiv g\tau \ll 1$ so energy exchange is $\mathcal{O}(\phi^2)$, is reset externally, and departs. After tracing out the outgoing pair and coarse-graining over many events, the cavity dynamics reduce to a linear birth-death (balance) equation in which the stream contributes an effective pump/drag term proportional to the collision strength $R\phi^2$ (with an additional detuning/filter factor if $\Delta \neq 0$),[6]

$$\dot{n} = -\kappa_{\text{eff}}(n - \bar{n}_{\text{bath}}) + R\phi^2[r_1 - (r_2 - r_1)n], \tag{1}$$

where $\phi = g\tau$, $R$ is the arrival rate, $r_1$ and $r_2$ encode the prepared up/down statistics of the incoming object, and

$$\bar{n}_{\text{bath}} = \frac{1}{e^{\hbar\omega_c/k_B T_{\text{bath}}} - 1}. \tag{2}$$

A persistent DQD, by contrast, remains coupled over many exchange events and can store correlations and energy; without re-preparation it need not act as a memoryless reservoir, and the joint cavity-DQD state matters. The bridge between these regimes is obtained by endowing the persistent DQD (single- or two-emitter) with a fast, Markovian re-preparation ("clamp/reset") that renders the emitter effectively memoryless on the cavity-evolution timescale. Concretely, the reset (or population-clamping) rate should be the fastest DQD timescale,

$$\gamma_{\text{res}} \gg \{\kappa, \Gamma_c(\Delta)\} \tag{2a}$$

so that the emitter populations remain near fixed values set by the reset channel. In the same weak-coupling/secular regime $g \ll \kappa + \gamma_\perp$, the cavity-emitter exchange is controlled by the Lorentzian spectral overlap with the effective energy-exchange rate

$$\Gamma_c(\Delta) = \frac{4g^2(\kappa + \gamma_\perp)}{(\kappa + \gamma_\perp)^2 + 4\Delta^2} \tag{3}$$

In this limit one can identify

$$R\phi^2 \leftrightarrow \Gamma_c(\Delta), \tag{4}$$

restoring a collision-like evolution for the cavity (with the same detuning filter). Away from this limit (slow re-preparation and/or finite intrinsic relaxation $\gamma_1$), persistent dynamics deviate from the stream model due to memory, saturation, and dependence on initial conditions; additional dephasing $\gamma_\phi$ (entering $\gamma_\perp$) further reduces spectral overlap and suppresses collective interference.

We adopt the following symbols and default values (used unless otherwise specified):

- $\kappa$:[‡] cavity damping rate to its environment (moderately weak tether), baseline $\kappa = 10\ kHz$.
- $g = g_\perp$: transverse Jaynes-Cummings coupling between the cavity mode and the active DQD transition; baseline $g/2\pi = 0.5\ MHz$.
- $\gamma_1$: DQD longitudinal relaxation (population decay) rate, $T_1^{-1}$, set by coupling to the phonon (and/or charge) environment and generally temperature dependent. It governs the decay of populations (diagonal density-matrix elements) in the energy eigenbasis.
- $\gamma_\phi$: DQD pure dephasing rate, $T_\phi^{-1}$, describing phase randomization without energy exchange. It contributes only to the decay of coherences (off-diagonal density-matrix elements).
- $\gamma_\perp$: homogeneous transverse decoherence (linewidth) rate, defined by $\gamma_\perp \equiv 1/T_2 = \gamma_1/2 + \gamma_\phi$. It sets the Lorentzian width entering spectral-overlap quantities (e.g., $\Gamma_c$). The factor 1/2 is the standard Bloch-equation result that population relaxation at rate $1/T_1$ contributes $1/(2T_1)$ to the decay of coherence ($1/T_2$).

---

[‡] Here $\kappa$ denotes the effective (loaded) energy damping rate of the cavity mode, i.e., the total linewidth including internal loss and external coupling to measurement/control ports ($\kappa = \kappa_{int} + \kappa_{ext}$), and any additional engineered loading channels if present.

- $\Delta$: detuning, $\Delta = \omega_{qd} - \omega_c$, where $\omega_{qd}$ is the active DQD transition frequency and $\omega_c$ is the cavity frequency.

- $\omega_c$: cavity angular frequency; baseline $\omega_c/2\pi = 5\ GHz$.

- $T_{\text{bath}}$: cryostat (phonon) bath temperature entering $\bar{n}_{\text{bath}}$; global-baseline studies may use $T_{\text{bath}} = 10\ K$ unless otherwise stated.

- $T_{DQD}$: effective reservoir temperature used for comparisons to the collision model

- $\tau, R$: collision-model parameters (reference only), with $\tau = 50\ ns$ and $R = 5 \times 10^6\ s^{-1}$.

With $\tau = 50\ ns$ and $g/2\pi = 0.5\ MHz$ we have $g \approx 2\pi \cdot 0.5\ MHz$ and

$$\phi = g\tau \approx 0.16, \tag{5}$$

which lies comfortably in the weak-collision regime. In the persistent DQD theory we work directly with $\Gamma_c(\Delta)$ in Eq. (3) and include $\gamma_\perp$ in the spectral overlap, recovering the collision-model structure through Eq. (4) in the fast-reset (or strong-dephasing) limit. Subsequent subsections detail the DQD Hamiltonian mapping to $g$ and $\Delta$, the Lindblad formulation with $\gamma_1$ and $\gamma_\phi$, and the two-emitter (DQD-both-dots-coupled) case that realizes correlation-assisted cooling analogous to the stream of internally correlated pairs.

## 3 Physical DQD-cQED model

### 3.1 Gate-defined DQD spin-Hamiltonian

We model a gate-defined DQD occupied by two electrons in the half-filling limit with each QD contain an electron, including spin-orbit coupling (SOC)-induced anisotropic exchange. The effective two-spin Hamiltonian (Heisenberg+Dzyaloshinskii-Moriya (DM)+symmetric anisotropy) in a uniform external field reads[7]

$$H_{\text{eff}} = \Delta_z (S_1^z + S_2^z) + J\,\mathbf{S}_1 \cdot \mathbf{S}_2 + \mathbf{D} \cdot (\mathbf{S}_1 \times \mathbf{S}_2) + \mathbf{S}_1 \cdot \overleftrightarrow{\Gamma} \cdot \mathbf{S}_2, \tag{6}$$

where $\Delta_z$ is the Zeeman splitting (including angular dependence and SOC renormalization), $J$ is the isotropic Heisenberg exchange coupling, $D$ is the DM interaction, and $\overleftrightarrow{\Gamma}$ denotes the anisotropy tensor. In the special case of the nanowire DQD model of Ref. [7], the SOC-modified Zeeman splitting can be written as $\Delta_z = g^*\mu_B B\, f(\vartheta, \varphi, x_{so})$, and the exchange parameters are set by the bare exchange $J_0$ and the geometry through $\zeta = 2d/x_{so}$ ($2d$ is the interdot separation and $x_{so}$ the spin-orbit length): $J = J_0 \cos^2\zeta$, $\mathbf{D} = J_0 \sin(2\zeta)\hat{v}$, and $\overleftrightarrow{\Gamma} = J_0 \sin^2\zeta\,(2\hat{v}\hat{v} - \mathbf{I})$, where $\hat{v}$ is the unit vector defining the DM (SOC) axis.

For the DM-axis rotation and inhomogeneous effective fields, following the standard Shekhtman rotation,[8,9] one can eliminate the DM term by opposite spin rotations about the DM axis (see Appendix), at the cost of a renormalized isotropic exchange $J_0$ and inhomogeneous local effective fields acting on the two spins:

$$H' = J_0\,\mathbf{S}_1 \cdot \mathbf{S}_2 + g^*\mu_B(\mathbf{B}_1 \cdot \mathbf{S}_1 + \mathbf{B}_2 \cdot \mathbf{S}_2), \tag{7}$$

where $J_0 \simeq \sqrt{J^2 + |\mathbf{D}|^2} + \mathcal{O}(\Gamma)$ and the uniform laboratory field $\mathbf{B}$ maps to unequal local fields $\mathbf{B}_{1,2}$ in the rotated frame. Here $\mathbf{B}_1 = R(\hat{n}, +\alpha)\,\mathbf{B}$, and $\mathbf{B}_2 = R(\hat{n}, -\alpha)\,\mathbf{B}$ are the rotated local effective fields (Appendix A.1-A.2), with the uniform laboratory field $B = |\mathbf{B}|$ entering $\Delta_z = g^*\mu_B B$ (or $\Delta_z = g^*\mu_B B f$ if we keep the SOC factor $f$). The inhomogeneity ($\mathbf{B}_1 \neq \mathbf{B}_2$) drives singlet-triplet mixing and underlies cavity-active transverse matrix elements in the DQD eigenbasis.

The matrix form of the Hamiltonian (7) is written in the singlet-triplet basis with the ordered basis $\{|T_+\rangle, |T_0\rangle, |T_-\rangle, |S_0\rangle\} = \{|\uparrow\uparrow\rangle, (|\uparrow\downarrow\rangle+|\downarrow\uparrow\rangle)/\sqrt{2}, |\downarrow\downarrow\rangle, (|\uparrow\downarrow\rangle-|\downarrow\uparrow\rangle)/\sqrt{2}\}$. In this representation, a minimal parametrization that captures the inhomogeneous-field-induced mixing is

$$H_{\text{eff}}^{(S/T)} = \begin{pmatrix} \frac{J_0}{4} + \delta & 0 & 0 & -ik \\ 0 & \frac{J_0}{4} & 0 & 0 \\ 0 & 0 & \frac{J_0}{4} - \delta & -ik \\ ik & 0 & ik & -\frac{3J_0}{4} \end{pmatrix}, \quad \delta = \Delta_z\sqrt{1-\beta_y^2}, \quad k = \frac{\beta_y \Delta_z}{\sqrt{2}}. \tag{8}$$

where the parameters are defined in the Appendix A.3. Here, $\beta_y \in [0,1]$ quantifies the transverse (y-like) inhomogeneous component produced by SOC/DM rotation. For $k \neq 0$, the Hamiltonian mixes $|S_0\rangle$ with $|T_\pm\rangle$, so the total $S_z$ is not conserved; however, $|T_0\rangle$ remains exactly decoupled at energy $\lambda_1 = J_0/4$. The remaining three eigenvalues are obtained by diagonalizing the 3×3 sub-block in the $\{|T_+\rangle, |T_-\rangle, |S_0\rangle\}$ sector. For the general case $k \neq 0$ and $\delta \neq 0$, this yields one analytic eigenvalue $\lambda_1 = J_0/4$ with eigenvector $|\psi_1\rangle = (0,1,0,0)^T$, and three additional eigenvalues $\lambda_{j+2}$ ($j = 0,1,2$) given by the cubic solution

$$\lambda_{j+2} = -\frac{J_0}{12} + 2\sqrt{\frac{-p}{3}} \cos\left(\frac{1}{3}\cos^{-1}\left(\frac{3q}{2p}\sqrt{\frac{-3}{p}}\right) - \frac{2\pi j}{3}\right), \quad j = 0, 1, 2$$

with corresponding normalized eigenvectors

$$|\psi_{j+2}\rangle = \frac{1}{N_{j+2}}\begin{pmatrix} A_{j+2} \\ 0 \\ B_{j+2} \\ 1 \end{pmatrix}, \quad N_{j+2} = \sqrt{1 + |A_{j+2}|^2 + |B_{j+2}|^2},$$

and

$$A_{j+2} = ik/\left(\frac{J_0}{4} + \delta - \lambda_{j+2}\right), \quad B_{j+2} = ik/\left(\frac{J_0}{4} - \delta - \lambda_{j+2}\right),$$

where

$$p = -\left(\frac{J_0^2}{3} + \delta^2 + 2k^2\right), \quad q = \frac{2J_0^3}{27} - \frac{2J_0\delta^2}{3} + \frac{2J_0 k^2}{3}.$$

In the special case $k = 0$ and $\delta = 0$, the Hamiltonian is diagonal and the eigenstates remain the bare singlet-triplet states. For small $k$ and $\delta$, one may alternatively use (degenerate) perturbation theory in the $\{|T_+\rangle, |T_-\rangle, |S_0\rangle\}$ manifold.

To connect this two-spin description to cavity coupling and single-dot populations/observables, we transform from the singlet-triplet basis to the product basis $\{|\uparrow\uparrow\rangle, |\uparrow\downarrow\rangle, |\downarrow\uparrow\rangle, |\downarrow\downarrow\rangle\}$ using a fixed unitary operator $U$ (given explicitly in the Appendix). Denoting the two-spin density matrix by $\rho_{\text{pair}}$ (see Appendix), the density matrix in the product basis is given by

$$\rho_{\text{prod}} = U \rho_{\text{pair}} U^\dagger, \tag{9}$$

For the single-emitter case, we obtain the excitation and ground populations of the active QD (QD A) from its reduced density matrix, $\rho_A = \text{Tr}_B[\rho_{\text{prod}}]$, where the partial trace is taken over the inactive dot (QD B).

In the rotated (DM-free) frame, the local quantization axis for the active dot is defined by $\hat{\mathbf{w}}_1 = \frac{\mathbf{B}_1}{|\mathbf{B}_1|}$. The probability that the QD A is excited along $\hat{\mathbf{w}}_1$ is then

$$p_{A,e} = \text{Tr}\left[\Pi_{A,e} \rho_A\right] = \frac{1}{2} - \langle \mathbf{S}_1 \cdot \hat{\mathbf{w}}_1 \rangle, \tag{10}$$

where $\mathbf{S} = \boldsymbol{\sigma}/2$ (dimensionless spin, eigenvalues $\pm 1/2$), and the operator $\Pi_{A,e} = (\mathbf{I} - \boldsymbol{\sigma}_1 \cdot \hat{\mathbf{w}}_1)/2$. The quantities that enter the cavity balance equations are the active-dot excitation and ground probabilities for the cavity-coupled transition, $r_1 = p_{A,e}$ and $r_2 = 1 - r_1$. Their explicit dependence on $(J_0, \delta, k)$ and on the prepared/reset setpoint state (thermal or otherwise) is derived in Sec. 5.3.

Remarks: (i) Eq. (7) provides a compact, device-independent representation for singlet-triplet mixing; the microscopic dependences on gate detuning $\varepsilon$, tunnel coupling $t_c$, spin-orbit length $x_{so}$, interdot spacing $2d$, and applied field $\mathbf{B}$ are summarized in the Appendix. (ii) For the two-emitter case (both dots coupled to the cavity), the same rotated-frame construction applies, but the cavity couples to both local quantization axes, leading naturally to collective bright/dark channels discussed in Sec. 6.

### 3.2 Cavity coupling channels and selection rules

The electric dipole interaction between a DQD and a single cavity mode of frequency $\omega_c$ decomposes, when expressed in the DQD eigenbasis, into transverse (energy-exchange) and longitudinal (dispersive) components. Denoting by $a$ ($a^\dagger$) the cavity annihilation (creation) operators and by $\sigma_\pm, \sigma_z$ the Pauli operators of the DQD defined along its local quantization axis, the generic interaction Hamiltonian can be written as

$$H_{\text{int}} = \hbar g_\perp \left( a\, \sigma_+ + a^\dagger \sigma_- \right) + \hbar g_\parallel \left( a + a^\dagger \right) \sigma_z \tag{11}$$

The transverse term describes excitation exchange between the DQD and the cavity, $|g\rangle \leftrightarrow |e\rangle$, and constitutes the dominant cooling channel. In contrast, the longitudinal term does not induce first-order energy exchange; instead, it produces ac-Stark shifts and contributes to dephasing.

The relative magnitudes of the transverse and longitudinal couplings are controlled by the DQD mixing angle $\theta$, which is set by the gate detuning $\varepsilon$ and the interdot tunnel coupling $t_c$ (see the Appendix). The mixing angle and the DQD level splitting are given by:

$$\tan\theta = \frac{2t_c}{\varepsilon}, \qquad \hbar\omega_{qd} = \sqrt{\varepsilon^2 + 4t_c^2}, \tag{12}$$

leading to the coupling strengths

$$g_\perp = g_0 \sin\theta, \quad g_\parallel = g_0 \cos\theta, \tag{13}$$

where $g_0$ is determined by the electric-dipole matrix element of the DQD and the cavity vacuum electric field. Biasing the device near the charge-degeneracy point ($\varepsilon \approx 0$, the "sweet spot") yields $\theta \approx \pi/2$, which results in $g_\perp \approx g_0$ and $g_\parallel \approx 0$. This maximizes the transverse coupling for energy exchange (e.g., cavity-assisted refrigeration), while suppressing longitudinal coupling and the associated photon-number-dependent dephasing of the DQD.

In the near-resonant, weak-coupling regime, with $g_\perp, g_\parallel \ll \omega_c, \omega_{qd}$, and within the rotating-wave approximation (RWA) applied to the exchange channel, the system Hamiltonian in the cavity frame takes the form

$$H = \hbar\omega_c\, a^\dagger a + \frac{\hbar\omega_{qd}}{2} \sigma_z + \hbar g_\perp \left( a\, \sigma_+ + a^\dagger \sigma_- \right) + \hbar g_\parallel \left( a + a^\dagger \right) \sigma_z. \tag{14}$$

We treat QD1 as the active quantum dot interacting with the cavity. In the rotated frame, its transition frequency is set by the local effective field $B_1$ via $\hbar\omega_{qd} = g^* \mu_B |B_1|$ (equivalently, $\omega_{qd} \propto B_1 \cdot \hat{w}_1$ with $\hat{w}_1 = B_1/|B_1|$). The resulting selection rules are determined by the local quantization axis $\hat{w}_1$ of the active dot. Only the component of the cavity electric field that is transverse to $\hat{w}_1$ contributes to $g_\perp$ and drives $|g\rangle \leftrightarrow |e\rangle$ transitions. The longitudinal component couples through $g_\parallel$ and primarily generates dispersive shifts and dephasing. In the remainder of this work, we operate close to $\varepsilon \approx 0$ in order to suppress $g_\parallel$ and focus on the transverse Jaynes-Cummings exchange governed by $g_\perp$ and the detuning $\Delta$.

## 4 Open quantum dynamics (Lindblad formulation)

We describe the DQD-cavity system using a thermal Lindblad master equation with independent, Markovian baths for the cavity mode and the DQD. Combining the effective DQD Hamiltonians in Eqs. (6)-(8) with the interaction in Eq. (11), and specializing to the RWA system Hamiltonian in Eq. (14), the density operator $\rho$ evolves according to:

$$\dot\rho = -\frac{i}{\hbar}[H,\rho] + \kappa^{(-)} \mathcal{D}[a]\rho + \kappa^{(+)} \mathcal{D}[a^\dagger]\rho + \gamma_\downarrow^{(\text{ph})} \mathcal{D}[\sigma_-]\rho + \gamma_\uparrow^{(\text{ph})} \mathcal{D}[\sigma_+]\rho + \frac{\gamma_\phi}{2} \mathcal{D}[\sigma_z]\rho, \tag{15}$$

where the Lindblad dissipator is defined as:

$$\mathcal{D}[L]\rho \equiv L\rho L^\dagger - \frac{1}{2}\{L^\dagger L, \rho\}. \tag{16}$$

The cavity mode is coupled to a thermal bosonic bath at temperature $T_{\text{bath}}$ with mean occupation

$$\bar{n}_{\text{bath}}(\omega_c, T_{\text{bath}}) = \frac{1}{\exp{(\hbar\omega_c/k_B T_{\text{bath}})}-1}, \tag{17}$$

In the standard thermal Lindblad form this implies emission and absorption rates $\kappa^{(-)}$ and $\kappa^{(+)}$ related to the bare energy-damping rate $\kappa$ by $\kappa^{(-)} = \kappa(\bar{n}_{\text{bath}} + 1), \kappa^{(+)} = \kappa \bar{n}_{\text{bath}}$ so that the bath drives the cavity toward $\bar{n}_{\text{bath}}$ in the absence of other couplings.

The DQD relaxes through coupling to a bosonic (phonon) environment characterized by the Bose occupation factor

$$n_B(\omega_{qd}, T_{\text{bath}}) = \frac{1}{\exp\left(\frac{\hbar\omega_{qd}}{k_B T_{\text{bath}}}\right)-1} \tag{18}$$

The corresponding temperature-dependent upward and downward transition rates are taken as

$$\gamma_\uparrow^{(\text{ph})} = \gamma_0 n_B, \quad \gamma_\downarrow^{(\text{ph})} = \gamma_0(n_B + 1), \quad \gamma_1 = \gamma_\uparrow^{(\text{ph})} + \gamma_\downarrow^{(\text{ph})}, \tag{19}$$

where $\gamma_0$ is the zero-temperature relaxation rate (i.e., the T $\to$ 0 limit of $\gamma_\downarrow^{ph}$). These rates satisfy the detailed balance condition by construction. Pure dephasing is included phenomenologically through the term $(\gamma_\phi/2)\mathcal{D}[\sigma_z]\rho$, and we define the homogeneous (Lorentzian) coherence linewidth

$$\gamma_\perp = \gamma_1/2 + \gamma_\phi. \tag{20}$$

This $\gamma_\perp$ is the linewidth that controls the decay of DQD coherences and enters spectral-overlap quantities such as the cavity-induced exchange rate $\Gamma_c$.

To capture slow charge noise or ensemble disorder, we optionally include an additional Gaussian inhomogeneous broadening $\sigma_{\text{inh}}$ of the DQD transition. In spectral-overlap expressions, e.g., the Lorentzian factors that appear in $\Gamma_c(\Delta)$, this can be incorporated by replacing the Lorentzian with a Voigt profile,

$$V(\Delta; \Gamma, \sigma_{\text{inh}}) = \text{Re}\left[w\left(\frac{\Delta + i\Gamma}{\sqrt{2}\,\sigma_{\text{inh}}}\right)\right]/(\sigma_{\text{inh}}\sqrt{2\pi}), \tag{21}$$

where $w(z)$ is the Faddeeva function. In practice, one may evaluate the Voigt profile numerically, or incorporate inhomogeneous broadening in rate expressions by replacing the Lorentzian factor with the Voigt profile and using an effective Voigt FWHM. Here the underlying Lorentzian half-width is $\Gamma = (\kappa + \gamma_\perp)/2$ (so the Lorentzian FWHM is $2\Gamma = \kappa + \gamma_\perp$). In all cases, $\gamma_\phi$ contributes only to coherence decay (pure dephasing), while $\gamma_1$ governs the temperature-dependent $T_1$ relaxation and satisfies detailed balance through Eq. (19).

## 5 Weak-coupling reduction and effective exchange

### 5.1 Spectral-overlap exchange rate

In the weak-coupling, near-resonant regime ($g \ll 2\Gamma$), where the RWA is valid and fast coherences can be secularized, the cavity and DQD act as mutually broadened Lorentzian reservoirs. Adiabatic elimination of the exchange coherences (equivalently, a Golden-rule evaluation from the convolution of the cavity and DQD spectra) yields a symmetric exchange rate governed by the sum of homogeneous widths $\Gamma_c(\Delta)$.

This Lorentzian overlap captures both Purcell-like exchange on resonance and detuning-induced suppression off resonance.

Off resonance, the DQD samples a detuning-filtered photon population. Defining $n \equiv \langle a^\dagger a \rangle$, the effective (spectrally overlapped) photon number entering the upward/downward exchange terms is

$$\tilde{n} = \frac{n}{1+(\Delta/\Gamma)^2}. \tag{22}$$

In subsequent rate equations, the cavity-induced excitation and relaxation channels for the DQD take the standard form $\Gamma_\uparrow^{(\text{cav})} = \Gamma_c(\Delta)\,\tilde{n}$ and $\Gamma_\downarrow^{(\text{cav})} = \Gamma_c(\Delta)\,(\tilde{n} + 1)$, while the reciprocal cavity terms use the same $\Gamma_c(\Delta)$ with $n$ (or $\tilde{n}$) as appropriate. These expressions formalize the intuitive picture that detuning reduces both the exchange rate and the effective photon drive by the same spectral-overlap factor set by $2\Gamma$.

## 5.2 Single persistent emitter (one active dot mapped to TLS)

In the weak-coupling regime, the cavity photon number $n \equiv \langle a^\dagger a \rangle$ and the excited-state population of the active dot A, $p_e = p_{A,e}$, obey coupled balance equations. The dynamics are driven by the detuning-dependent cavity-QD exchange rate $\Gamma_c(\Delta)$ [Eq. (3)], with bath loading set by $\kappa$ and by the DQD phonon relaxation rate $\gamma_1$. For general detuning $\Delta$ the rate equations can be written as

$$\dot{n} = -\kappa(n - \bar{n}_{\text{bath}}) + \Gamma_c(\Delta)S, \tag{23}$$

$$\dot{p}_e = -\Gamma_c(\Delta)S - \gamma_1(p_e - p_{\text{th}}), \tag{24}$$

where $p_{\text{th}} = n_B/(2n_B + 1)$ is the thermal excited-state probability of the dot at the bath temperature, and the exchange bias is $S \equiv (n+1)p_e - n(1-p_e) = (2n+1)p_e - n$.

Off resonance, spectral overlap can be emphasized by replacing $n \to \tilde{n}$ from Eq. (22) inside $S$, i.e., $S \to (2\tilde{n} + 1)p_e - \tilde{n}$, which accounts for the detuning-filtered exchange seen by the dot-cavity channel.

To obtain a closed-form steady state, we set $\Delta = 0$ so that $\Gamma_c \equiv \Gamma_c(0)$ from Eq. (3). At stationarity, Eqs. (23) and (24) imply:

$$\Gamma_c S = \kappa(n - \bar{n}_{\text{bath}}) = -\gamma_1(p_e - p_{\text{th}}), \tag{25}$$

which yields the linear relation:

$$p_e = p_{\text{th}} - \frac{\kappa}{\gamma_1}(n - \bar{n}_{\text{bath}}). \tag{26}$$

Eliminating $p_e$ via Eqs. (23) and (24) produces a quadratic equation for the steady-state cavity population $n^*$:

$$a\, n^{*2} + b\, n^* + c = 0, \tag{27}$$

with coefficients:

$$\begin{aligned}
a &= -2\Gamma_c \kappa, \\
b &= \Gamma_c \gamma_1 (2p_{\text{th}} - 1) + \Gamma_c \kappa(2\bar{n}_{\text{bath}} - 1) - \kappa\gamma_1, \\
c &= \Gamma_c \gamma_1 p_{\text{th}} + \Gamma_c \kappa \bar{n}_{\text{bath}} + \kappa \gamma_1 \bar{n}_{\text{bath}}.
\end{aligned}$$

Since $a < 0$, the physical root is:

$$n^* = \frac{-b + \sqrt{b^2 - 4ac}}{2a}. \tag{28}$$

The corresponding DQD steady-state excitation probability follows from Eq. (26),

$$p_e^* = p_{\text{th}} - \frac{\kappa}{\gamma_1}(n^* - \bar{n}_{\text{bath}}). \tag{29}$$

Finally, effective temperatures are assigned by inverting the Bose and two-level detailed-balance relations,

$$T_{\text{cav}} = \frac{\hbar\omega_c}{k_B \ln(1 + 1/n^*)}, \tag{30}$$

$$T_{\text{DQD}} = \frac{\hbar\omega_{qd}}{k_B \ln(1/p_e^* - 1)}, \tag{31}$$

and

$$\frac{T_{\text{cav}}}{T_{\text{DQD}}} = \frac{\omega_c}{\omega_{qd}} \frac{\ln(1/p_e^* - 1)}{\ln(1 + 1/n^*)} \tag{33a}$$

Setting aside special operating points, it is useful to record several limiting cases that serve as sanity checks on Eqs. (22)-(33) and clarify how the coupled cavity-DQD dynamics interpolate between conserved-manifold behavior and bath-pinned equilibration.

If both dissipative channels are removed ($\kappa = \gamma_1 = 0$), Eq. (25) reduces to $\Gamma_c S = 0$, which implies

$$(n+1)p_e = n(1-p_e) \Rightarrow p_e = \frac{n}{2n+1}. \tag{32}$$

In this limit, the coupled system relaxes onto a conserved-excitation (memory) manifold rather than a unique thermal fixed point, and the long-time values depend explicitly on the initial conditions.

At the opposite extreme, if an engineered reset (or sufficiently strong dephasing) clamps the DQD excitation probability to a fixed value $p_e \approx r_1$ on timescales much shorter than $\Gamma_c^{-1}$ and $\kappa^{-1}$, then Eq. (23) reduces to a single closed equation for the cavity population,

$$\dot{n} = -\kappa(n - \bar{n}_{\text{bath}}) + \Gamma_c[r_1 - (r_2 - r_1)n], \tag{33}$$

which is precisely the collision-model form, with the identification $R\phi^2 \leftrightarrow \Gamma_c(\Delta)$ at the same detuning and linewidth. In this sense, the exchange channel plays the role of an effective stream-induced rate.

Two additional limits clarify the bath-pinned regimes. When $\gamma_1 \gg \Gamma_c$, the DQD rapidly equilibrates to its thermal value $p_e^* \to p_{th}(T_{bath}, \omega_{qd})$. In this regime the cavity is effectively bath-loaded, so $n^* \to \bar{n}_{bath}(T_{bath}, \omega_c)$, with only a small back-action correction controlled by $\Gamma_c/\kappa$. With the effective-temperature definitions in Eqs. (30)-(31), both subsystems are therefore pinned to the bath, $T_{DQD} \to T_{bath}$ and $T_{cav} \to T_{bath}$, implying $T_{cav}/T_{DQD} \to 1$. The "classical" versus "quantum" character is instead captured by the bath occupation: for $\hbar\omega_c \ll k_B T_{bath}$ one has $\bar{n}_{bath} \gg 1$ (classical regime), while for $\hbar\omega_c \gg k_B T_{bath}$ one has $\bar{n}_{bath} \ll 1$ (quantum regime), reflecting the suppression of thermal photons at large $\omega_c$.

Conversely, when $\kappa \gg \Gamma_c$, the cavity is pinned to the bath, $n^* \to \bar{n}_{\text{bath}}$, while the DQD relaxes toward $p_{th}$ with a small correction controlled by $\Gamma_c/\gamma_1$.

Finally, for finite detuning $\Delta \neq 0$, one replaces $\Gamma_c \to \Gamma_c(\Delta)$ and, if desired, substitutes the detuning-filtered photon number $n \to \tilde{n}$ from Eq. (22) inside the exchange bias $S$ to make the role of spectral filtering explicit. The steady-state algebra then proceeds analogously and yields the same quadratic structure as on resonance, with $\Gamma_c$ replaced by its detuned value $\Gamma_c(\Delta)$.

### 5.3 How the realistic DQD Hamiltonian sets the "reservoir statistics"

A realistic DQD fixes the effective "reservoir statistics" through its thermal spin state and the local quantization axis of the active dot. Starting from the DQD Hamiltonian in Eqs. (6)-(8), the two-spin system is first thermalized at temperature $T_{\text{bath}}$ and then reduced to the active dot.

The thermal state of the spin pair is constructed from $H_{eff}$ and written in the singlet-triplet basis as $\rho_{ST}$ (Appendix A.4), then mapped to the product basis and reduced to the active dot A by partial trace. Throughout, we define the active-dot excited state as the spin-down state along the chosen local quantization axis, $|e\rangle_A \equiv |\downarrow\rangle_A$, and the ground state as $|g\rangle_A \equiv |\uparrow\rangle_A$. Accordingly, for a general local axis $\hat{\mathbf{w}}_1$ the excited-state projector is $\Pi_{A,e} = (\mathbf{I} - \boldsymbol{\sigma}_1 \cdot \hat{\mathbf{w}}_1)/2$, so $p_{A,e} = \text{Tr}[\Pi_{A,e}\rho_A] = 1/2 - \langle \mathbf{S}_1 \cdot \hat{\mathbf{w}}_1 \rangle$. In the common case $\hat{\mathbf{w}}_1 = \hat{\mathbf{z}}$, and using the ordered S/T basis ($|T+\rangle, |T_0\rangle, |T_-\rangle, |S_0\rangle$), the marginal simplifies to $p_{A,e} = \rho_{33} + (\rho_{22} + \rho_{44})/2$, i.e., the probability that dot A is in $|\downarrow\rangle$. (In the minimal model of Eq. (8), $|T_0\rangle$ is decoupled, so no $T_0 - S_0$ coherence contributes to this marginal.) We then define the effective single-emitter "reservoir statistics" as

$$r_1 \equiv p_{A,e}, \quad r_2 \equiv 1 - r_1 \tag{34}$$

These coefficients $(r_1, r_2)$ constitute the microscopic inputs to the one-active-dot cooling theory in the refreshed-reservoir limit, where the emitter is rapidly reset (or clamped) between exchange events so that each interaction samples the same prepared distribution.

When an auxiliary, Markovian reset channel clamps the active-dot population on a timescale short compared to the exchange and cavity-leakage dynamics, the persistent DQD effectively emulates a memoryless stream. This is modeled by augmenting the master equation (Sec. 4) with

$$\mathcal{L}_{\text{res}}\rho = \Gamma_{\downarrow}^{(\text{res})} \mathcal{D}[\sigma_-]\rho + \Gamma_{\uparrow}^{(\text{res})} \mathcal{D}[\sigma_+]\rho + \frac{\gamma_{\phi,\text{res}}}{2} \mathcal{D}[\sigma_z]\rho, \tag{35}$$

with reset rates chosen to reproduce the DQD-derived statistics,

$$\Gamma_{\uparrow}^{(\text{res})} = \gamma_{1,\text{res}} r_1, \quad \Gamma_{\downarrow}^{(\text{res})} = \gamma_{1,\text{res}} r_2, \quad r_1 + r_2 = 1. \tag{36}$$

In the fast-reset (or strong-dephasing) limit,

$$\max\{\gamma_{1,\text{res}}, \gamma_{\phi,\text{res}}\} \gg \{\kappa, \Gamma_c(\Delta)\}, \tag{37}$$

the emitter population is effectively clamped, $p_e(t) \approx r_1$, and the cavity dynamics reduce to the collision-model form (cf. Eq. (33)),

$$\dot{n} = -\kappa (n - \bar{n}_{\text{bath}}) + \Gamma_c(\Delta) [r_1 - (r_2 - r_1) n]. \tag{38}$$

With the identification $R\phi^2 \leftrightarrow \Gamma_c(\Delta)$, the realistic DQD Hamiltonian determines the reservoir coefficients $r_1$ and $r_2$ through Eqs. (34)-(35), while the engineered reset enforces these as fixed statistics driving the cavity. The exchange bandwidth is set by $\Gamma_c(\Delta)$ in Eq. (3), governed by the spectral overlap width $\kappa + \gamma_\perp$.

It should be noted that the quantities $\gamma_{1,res}$ and $\gamma_{\phi,res}$ are microscopic Lindblad rates describing population reset and pure dephasing of the DQD, respectively, and appear explicitly in the master equation (Eq. 35). In Sec. 12.1 we introduce the composite parameter $\Gamma_{reset}$ to characterize the overall speed of the reset/clamp pathway at the level of the thermodynamic description. Throughout the manuscript, $\Gamma_{reset}$ should be understood as being of order $max(\gamma_{1,res}, \gamma_{\phi,res})$. The fast-reset condition $\Gamma_{reset} \gg (\kappa, \Gamma_c(\Delta))$ is therefore equivalent to Eq. (37).

## 6 Two persistent emitters (two-dot active case; Tavis-Cummings)

We now consider the case in which both dots are "active" and couple to the same cavity mode, forming a two-emitter Tavis-Cummings system (with optional inter-dot exchange). Relative to a single active dot, the always-on two-emitter configuration can access collective bright/dark interference that effectively increases the exchange bandwidth available for refrigeration, but the enhancement is fragile to frequency mismatch and dephasing.

A minimal model is

$$H = \hbar\omega_c a^\dagger a + \sum_{j=1}^{2}\left[\frac{\hbar\omega_{qd,j}}{2}\sigma_z^{(j)} + \hbar g_j(a\sigma_+^{(j)} + a^\dagger\sigma_-^{(j)})\right] + \hbar\lambda(\sigma_+^{(1)}\sigma_-^{(2)} + \sigma_-^{(1)}\sigma_+^{(2)}), \tag{39}$$

where $\omega_{qd,j}$ are the local two-level splittings of dot $j$ in the rotated (DM-free) frame, and the Pauli operators $\sigma_z^j, \sigma_\pm^j$ are defined in the local eigenbasis of that dot. Specifically, the local quantization axis is $\hat{w}_j \equiv B_j/|B_j|$, so $\sigma_z^j \equiv \sigma^j \cdot \hat{w}_j$ and $\sigma_\pm^j$ are the corresponding raising/lowering operators with respect to $\hat{w}_j$. The Zeeman term is therefore diagonal in this basis, $H_Z^j = (\hbar\omega_{qd,j}/2)\sigma_z^j$, with $\hbar\omega_{qd,j} = \Delta_z(|B_j|/B)$, while $g_j$ are the transverse cavity-DQD matrix elements (maximized near the charge sweet spot), and $\lambda$ is the tunable inter-dot exchange coupling. We define the detunings

$$\Delta_j = \omega_{qd,j} - \omega_c, \quad \Delta_{12} = \omega_{qd,1} - \omega_{qd,2}. \tag{40}$$

A more general inter-dot interaction may also include a longitudinal term $(\hbar\chi/4)\sigma_z^{(1)}\sigma_z^{(2)}$ arising from the $S_1^z S_2^z$ component of $J_0 \, S_1 \cdot S_2$; in the single-excitation (bright/dark) subspace this contributes only a constant shift and is therefore absorbed into the definition of the transition frequencies $\omega_{qd,j}$ (or dropped). We retain the flip-flop term $\lambda$, which controls bright/dark hybridization and the collective spectral overlap.

In the symmetric limit of equal frequencies and equal coupling phases, it is convenient to introduce collective ladder operators

$$S_+^B = \frac{\sigma_+^{(1)} + \sigma_+^{(2)}}{\sqrt{2}}, \quad S_+^D = \frac{\sigma_+^{(1)} - \sigma_+^{(2)}}{\sqrt{2}}, \quad S_-^{B,D} = (S_+^{B,D})^\dagger, \tag{41}$$

which generate bright and dark modes, respectively. In this basis, the cavity couples exclusively to the bright mode:

$$H_{\text{int}} = \hbar g_B(a S_+^B + a^\dagger S_-^B), \quad g_B = |g_1 + g_2|/\sqrt{2}, \tag{42}$$

while the dark mode ($S^D$) is decoupled in the perfectly symmetric limit. Small frequency mismatch ($\Delta_{12} \neq 0$), unequal phases, or finite inter-dot exchange $\lambda$ admix bright and dark character, giving the nominally dark mode a weak effective coupling and partially lifting the "darkness."

In the weak-coupling, secular regime, the bright-mode exchange rate follows the same detuning-filtered (Lorentzian spectral-overlap) structure as Eq. (3), with g replaced by $g_B$ and the homogeneous linewidth replaced by an average transverse rate $\bar{\gamma}_\perp = (\gamma_{\perp,1} + \gamma_{\perp,2})/2$. The collective exchange rate is then

$$\Gamma_{c,B}(\Delta_B) = \frac{2\, g_B^2\, \bar{\Gamma}}{\bar{\Gamma}^2 + \Delta_B^2}, \quad \Delta_B \approx \tfrac{1}{2}(\Delta_1 + \Delta_2), \quad 2\bar{\Gamma} = \kappa + \bar{\gamma}_\perp \tag{43}$$

For $g_1 = g_2 = g$ and exact symmetry, $g_B = \sqrt{2}g$, so $\Gamma_{c,B}$ is enhanced by a factor of two relative to a single emitter (at fixed $\kappa$ and $\bar{\gamma}_\perp$). This collective gain is fragile to frequency mismatch $|\Delta_{12}|$, which detunes the two QD spins and admixes dark and bright states, and to dephasing $\gamma_{\phi,j}$, which spoils phase coherence. A simple estimate of this degradation is obtained by broadening the effective bright-channel linewidth in Eq. (43), i.e., $\bar{\gamma}_\perp \to \bar{\gamma}_\perp + |\Delta_{12}|$ (equivalently $2\bar{\Gamma} \to 2\bar{\Gamma} + |\Delta_{12}|$), capturing the rapid collapse of the collective enhancement as mismatch exceeds the homogeneous width. When emitter-emitter coherences are strongly suppressed (by large dephasing or large mismatch), the dynamics are well approximated by two independent exchange channels,

$$\dot{n} = -\kappa(n - \bar{n}_{\text{bath}}) + \sum_{j=1}^{2} \Gamma_{c,j}(\Delta_j) S^{(j)}, \tag{44}$$

$$\dot{p}_e^{(j)} = -\Gamma_{c,j}(\Delta_j) S^{(j)} - \gamma_{1,j}\left(p_e^{(j)} - p_{\text{th}}^{(j)}\right), \tag{45}$$

with $p_{\text{th}}^{(j)} = n_B^{(j)}/(2n_B^{(j)} + 1)$, the exchange biases $S^{(j)} = (n+1)p_e^{(j)} - n(1 - p_e^{(j)})$, and single-emitter spectral overlaps

$$\Gamma_{c,j}(\Delta_j) = \frac{2g_j^2}{\Gamma_j^2 + \Delta_j^2}, \quad 2\Gamma_j = \kappa + \gamma_{\perp,j}, \quad \gamma_{\perp,j} = \gamma_{1,j}/2 + \gamma_{\phi,j}. \tag{46}$$

In the symmetric, coherent limit, these independent-emitter equations reduce to the collective description, with the net exchange governed by the bright channel and $\Gamma_{c,B}$ replacing the simple sum $\sum_j \Gamma_{c,j}$. Operationally, the collective enhancement opens a correlation-assisted cooling window: for fixed $\bar{\Gamma} = (\kappa + \bar{\gamma}_\perp)/2$, increasing $g_B$ deepens refrigeration provided that the accompanying mismatch and dephasing do not broaden the collective response enough to offset the gain. As $\gamma_{\phi,j}$ or $|\Delta_{12}|$ increases, the enhancement collapses continuously toward the independent-emitter regime.

The inter-dot exchange term $\lambda$ further mixes the single-excitation manifold (in the rotated DM-free frame, $\lambda$ is set by the underlying exchange scale, $\lambda \approx J_0/(2\hbar)$). When $|\lambda|$ is small, it can partially brighten nominally dark superpositions and thereby aid equilibration. When $2|\lambda|$ becomes comparable to the homogeneous width $2\bar{\Gamma}$, it resolves a bright-dark (or bonding-antibonding) splitting and reduces spectral overlap with the cavity at $\Delta_B \approx 0$. Optimal cooling therefore occurs when $\lambda$ preserves an effectively single bright resonance near $\Delta_B \approx 0$ without excessive splitting, thereby maximizing $\Gamma_{c,B}$.

## 7   Comparison with the collision model

This section relates the persistent DQD framework (Secs. 3-5) to the collision model based on a Poisson stream of internally correlated pairs.[6] The two descriptions become equivalent when the persistent emitter behaves as a memoryless reservoir, and they diverge when the emitter stores correlations or energy.

If the DQD is rendered effectively memoryless, either by an auxiliary reset channel or by sufficiently strong clamping through dissipation, its influence on the cavity reduces to the same Markovian pump/drag structure as in the collision (stream) model, with the spectral-overlap exchange rate playing the role of the stream-induced rate.

For a single active dot, the reduction occurs when the emitter population is clamped on a timescale short compared to both exchange and cavity leakage. Concretely, in the regime $\gamma_{\text{res}}, \gamma_{\phi,\text{res}} \gg \{\kappa, \Gamma_c(\Delta)\}$, the emitter tracks a fixed value $p_e(t) \approx r_1$, and the cavity equation reduces to Eq. (38) with the identification $R\phi^2 \leftrightarrow \Gamma_c(\Delta)$, where $\Gamma_c(\Delta)$ is given by Eq. (3). A second, "bath-clamp" limit arises when $\gamma_1 \gg \Gamma_c(\Delta)$, in which case the emitter follows the thermal population $p_e \to p_{th}$, so the effective stream coefficients become $r_1 = p_{th}, r_2 = 1 - p_{th}$.

In both cases, the cavity experiences a Markovian pump/drag whose strength is set by the same detuning-filtered exchange rate $\Gamma_c(\Delta)$.

For two active dots, the same logic applies but with the collective bright channel replacing the single-emitter exchange. In the symmetric, coherent limit the cavity couples predominantly to the bright superposition with $g_B$, and the relevant exchange rate is $\Gamma_{c,B}(\Delta_B)$ from Eq. (43). With fast collective reset (or population clamping that maintains the intended bright-mode occupancy while preventing memory buildup), the mapping generalizes to $R\phi^2 \leftrightarrow \Gamma_{c,B}(\Delta_B)$.

Small frequency mismatch and dephasing are tolerable only insofar as they do not strongly admix the dark mode or fragment the bright resonance, since either effect reduces the effective exchange bandwidth available to cool the cavity.

When reset is absent or slow, the persistent-emitter implementation deviates from the stream model in ways that are intrinsic to solid-state reservoirs. First, the emitter stores correlations and energy; in the limiting case $\kappa = \gamma_1 = 0$, Eq. (32) shows that the system relaxes onto a conserved-excitation manifold, $(n+1)p_e = n(1-p_e)$, so long-time values depend on initial conditions rather than converging to a unique fixed point. Second, the exchange bias $S = (2n+1)p_e - n$ saturates as $p_e$ approaches its bounds $0 \le p_e \le 1$, limiting the benefit of increasing $\Gamma_c$; the collision model avoids this saturation by continually refreshing the reservoir state. Third, in the two-emitter geometry, bright-dark interference produces $\Gamma_{c,B} \propto g_B^2$, but the enhancement is degraded by frequency mismatch $|\Delta_{12}|$ and by dephasing $\gamma_{\phi,j}$, both of which destroy the coherence required to maintain a protected dark sector and a well-defined bright channel. Population trapping in dark subspaces has no analogue in the ideal collision picture, where correlations are externally prepared and reset between events. Finally, spectral suppression at large detuning $|\Delta| \gg 2\Gamma$ (or strong inhomogeneous broadening) reduces both $\Gamma_c(\Delta)$ and the filtered photon number $\tilde{n}$ from Eq. (22), weakening any mapping to an effective stream.

These considerations can be summarized as a qualitative regime chart. Stream-like single-emitter behavior requires (i) fast clamping, $\gamma_{1,res}$ (and/or $\gamma_{\phi,res}$) $\gtrsim O(1) \times max\{\kappa, \Gamma_c(\Delta)\}$, (a practical rule of thumb is a few times larger), and (ii) spectral overlap, $|\Delta| \lesssim 2\Gamma$, with $r_1$ and $r_2$ set by the microscopic construction in Sec. 5.3. Stream-like two-emitter behavior further requires approximate symmetry $g_1 \approx g_2, \Delta_{12} \approx 0$, so that $g_B$ is maximized, negligible leakage into the dark sector, and population preparation that maintains bright mode occupancy so that $\Gamma_{c,B}(\Delta_B)$ governs the exchange. Persistent, memory-dominated dynamics appear when $\Gamma_c(\Delta) \gtrsim \{\kappa, \gamma_1\}$ in the absence of reset, allowing emitter-cavity correlations to accumulate; in the two-emitter case this regime also admits dark-state trapping. The bath-pinned limits are recovered when $\kappa \gg \Gamma_c(\Delta)$ (cavity pinned, $n \to \tilde{n}_{bath}$), or $\gamma_1 \gg \Gamma_c(\Delta)$ (emitter pinned, $p_e \to p_{th}$), and the dephasing-/mismatch-dominated limit collapses the two-emitter dynamics to an incoherent sum of independent exchange channels described by Eqs. (44)-(46).

The key takeaway is that a persistent DQD reproduces collision-model physics when the emitter is made effectively memoryless (fast reset or strong population clamping) and when spectral overlap is maintained, $|\Delta| \lesssim 2\Gamma$.

Deviations arise from stored correlations, dark-state formation, and spectral mismatch, features absent in an ideal Poisson stream but central to realistic solid-state implementations. Designing for the stream-like regime therefore reduces to three practical requirements: maximize spectral overlap, enforce rapid clamping of emitter populations, and, in the two-emitter configuration, preserve bright mode coherence while suppressing leakage into dark subspaces.

## 8    Analytical results and scaling laws

We collect closed-form bounds and scaling relations for the steady-state cavity temperature $T_{cav}$ in terms of the key parameters $\{g, \kappa, \gamma_\perp, \gamma_1, \Delta\}$. Results are summarized for a persistent DQD both with and without an engineered reset; off resonance, one replaces $\Gamma_c \to \Gamma_c(\Delta)$ using Eq. (3).

In the clamped (stream-like) limit, an engineered reset (or sufficiently strong dephasing) pins the active-dot population to a fixed value determined by the microscopic single-dot statistics (Sec. 5.3). In this case, the cavity steady state can be expressed as a weighted mixture of two "setpoints": the bath occupation $\bar{n}_{\text{bath}}$ and a DQD-imposed occupation $n_{\text{set}}$ defined by the up/down statistics $(r_1, r_2)$. Specifically, we define

$$n_{\text{set}} \equiv \frac{r_1}{r_2 - r_1} = \frac{1}{e^{\frac{\hbar\omega_{qd}}{k_B T_{\text{set}}}} - 1}, \quad \frac{r_1}{r_2} = e^{-\hbar\omega_{qd}/k_B T_{\text{set}}}. \tag{47}$$

The steady photon number is then

$$n^*(\Delta) = \frac{\kappa \bar{n}_{\text{bath}} + \Gamma_c(\Delta) n_{\text{set}}}{\kappa + \Gamma_c(\Delta)}, \quad T_{\text{cav}} = \frac{\hbar \omega_c}{k_B \ln(1 + 1/n^*)}. \tag{48}$$

Cooling ($T_{\text{cav}} < T_{\text{bath}}$) occurs if and only if $n_{\text{set}} < \bar{n}_{\text{bath}}$, equivalently $T_{\text{set}} < T_{\text{bath}}$, and for any fixed detuning $\Delta$ the cavity temperature is bounded by

$$T_{\text{set}} \leq T_{\text{cav}} \leq T_{\text{bath}} \tag{49}$$

The lower bound is approached when $\kappa \ll \Gamma_c(\Delta)$, while $T_{\text{cav}} \to T_{\text{bath}}$ when $\Gamma_c(\Delta) \ll \kappa$.

Without clamp, the on-resonance steady state is given by the exact solution of Sec. 5.2: $n^*$ follows from Eq. (28) and $p_e^*$ from Eq. (29). As $\Gamma_c$ increases, $T_{\text{cav}}$ decreases from $T_{\text{bath}}$ toward a floor set by the DQD's own rethermalization through $p_{\text{th}}$ and by the approach to the exchange bias manifold $S = 0$. In the strong-exchange limit $\Gamma_c \gg \{\kappa, \gamma_1\}$, one finds to leading order

$$(2n^* + 1) p_e^* \approx n^*, \quad p_e^* \approx p_{\text{th}} - \frac{\kappa}{\gamma_1}(n^* - \bar{n}_{\text{bath}}), \tag{50}$$

so that the cavity approaches an effective setpoint implied by $p_{\text{th}}$, with corrections controlled by $\kappa/\gamma_1$.

These expressions highlight the main operating levers through $\Gamma_c(\Delta)$ and the spectral-overlap window. To maximize transverse exchange and overlap, one biases near the charge sweet spot $\varepsilon \approx 0$ to maximize $g_\perp$ and suppress $g_\parallel$ (Eqs. (12)-(13)), thereby increasing $\Gamma_c$ while reducing sensitivity to longitudinal noise. For a fixed linewidth $2\Gamma$, the exchange channel is largest for near-resonant detuning $|\Delta| \lesssim 2\Gamma$, which maximizes both $\Gamma_c(\Delta)$ and the filtered occupation entering Eq. (22). On resonance, the exchange rate scales as $\Gamma_c(0) = \frac{2g^2}{\Gamma}$.

Reducing $\gamma_\phi$ (hence $\gamma_\perp$) through sweet-spot operation and materials optimization therefore directly increases $\Gamma_c$; when $\gamma_\perp \gg \kappa$ the exchange is dephasing-limited, $\Gamma_c(0) \approx 4g^2/\gamma_\perp$.

Likewise, reducing $\kappa$ lowers bath loading and increases $\Gamma_c(0)$, allowing the engineered exchange channel to dominate the steady state.

Several limiting forms are then immediate. In the exchange-dominated regime $\Gamma_c \gg \kappa$ (with finite $\gamma_1$), Eq. (50) shows that the cavity approaches the DQD setpoint,

$$n^* \approx (\Gamma_c n_{\text{set}} + \kappa \bar{n}_{\text{bath}})/(\Gamma_c + \kappa), \Rightarrow T_{\text{cav}} \to T_{\text{set}} \text{ (clamped) or } T_{\text{cav}} \to T_{\text{DQD}} \text{ (unclamped) as } \Gamma_c/\kappa \to \infty \tag{51}$$

In the opposite emitter-pinned limit $\gamma_1 \gg \Gamma_c$, the DQD remains near its thermal population $p_{\text{th}}$ and the cavity simply mixes the bath occupation with the corresponding thermal setpoint implied by $p_{\text{th}}$ through Eq. (52):

$$p_e^* \to p_{\text{th}}, \quad n^*(\Delta) \approx \frac{\kappa \bar{n}_{\text{bath}} + \Gamma_c(\Delta) n_{\text{set}}(p_{\text{th}})}{\kappa + \Gamma_c(\Delta)}, \tag{52}$$

In the bath-dominated limit $\kappa \gg \Gamma_c$, one recovers

$$n^* \to \bar{n}_{\text{bath}}, T_{\text{cav}} \to T_{\text{bath}}, \text{ (weak back-action).} \tag{53}$$

with only weak back-action. In the emitter-pinned limit $\gamma_1 \gg \Gamma_c(\Delta)$, $p_e^* \to p_{th}$, so the "effective DQD temperature" inferred from Eq. (31) is simply $T_{\text{DQD}} \to T_{\text{bath}}$ ($p_{th}$ satisfies detailed balance at $\omega_{qd}$ and $T_{\text{bath}}$). An explicit expression for the temperature ratio then follows by combining Eq. (30) with $T_{\text{DQD}} = T_{\text{bath}}$: $T_{\text{cav}}/T_{\text{DQD}} = T_{\text{cav}}/T_{\text{bath}} = [\hbar\omega_c/(k_B T_{\text{bath}})]/\ln(1 + 1/n^*(\Delta))$, with $n^*(\Delta)$ given by Eq. (52). In the bath-dominated limit $\kappa \gg \Gamma_c(\Delta)$, Eq. (53) gives $n^* \to \bar{n}_{bath}$ and thus $T_{\text{cav}} \to T_{\text{bath}}$, while $p_e^*$ remains $p_{th}$ so $T_{DQD} \to T_{bath}$; therefore $T_{cav}/T_{DQD} \to 1$. If desired, the "equivalent bosonic setpoint" associated with a thermal TLS population can be written as $n_{set}(p_{th}) = p_{th}/(1 - 2p_{th})$ (TLS ↔ boson).

Off resonance, both $\Gamma_c(\Delta)$ and the filtered occupation $\tilde{n}$ are suppressed by the detuning factor $[1 + (\Delta/\Gamma)^2]^{-1}$, so that refrigeration is confined to the overlap window $|\Delta| \lesssim 2\Gamma$. Finally, for fixed $\{g, \kappa, \gamma_\perp, \gamma_1\}$, the minimum $T_{\text{cav}}$ is achieved near $\Delta \to 0$, while $\kappa \to 0$ pushes the clamped solution toward $T_{\text{set}}$ and the unclamped solution toward the $p_{\text{th}}$-implied manifold (Eq. (50)).

**Design implication.** For fixed $T_{\text{bath}}$, the dominant levers lowering $T_{\text{cav}}$ are (i) increasing the transverse coupling $g_\perp$, (ii) reducing $\kappa$ and $\gamma_\perp$ so that the exchange channel can compete with bath loading, and (iii) operating within the near-resonant overlap window. The absolute floor is set by the DQD-imposed setpoint $T_{\text{set}}$ in the clamped limit (or by the $p_{\text{th}}$-determined analogue in the persistent limit), while finite $\kappa$ and $\gamma_\perp$ determine how closely the cavity can approach that floor in practice.

## 9  Numerical methodology

We use two complementary routes to compute steady-state observables and parameter sweeps: (i) full Lindblad evolution of the coupled DQD-cavity density matrix, and (ii) reduced rate equations in the weak-coupling, secular regime. In the numerical calculations, $T_{\text{bath}}$ is the cavity-bath temperature; $T_{\text{set}}$ is the engineered reset setpoint that fixes the prepared populations $(r_1, r_2)$ in the clamped limit; and $T_{\text{DQD}}$ is an effective temperature inferred from the persistent emitter's steady-state population using the inversion relation in Eq. (31). In the collision-style comparisons, the prepared reservoir temperature used to set $(r_1, r_2)$ is taken to be $T_{set}$, and in the persistent no-clamp examples we assume $T_{\text{DQD}} = T_{\text{set}}$ (ideal active reset/algorithmic cooling). We label this common reference temperature as $T_{dot}$ in the plots for uniformity, so the y-axis uses the same prepared/reset temperature scale when comparing collision-like and clamped persistent results, even though the physical origin of that temperature differs between the two pictures.

Given device parameters and material constants, we construct the effective two-spin Hamiltonian using Eqs. (6)-(8) and the mapping steps summarized in the Appendix. From the resulting prepared (or thermal) two-spin state, we obtain the single-transition statistics relevant to the cavity through Eq. (34) (and the associated local-axis definitions introduced around Eq. (10)), which fixes $(r_1, r_2)$ for the clamped calculations via Eq. (36). The cavity coupling matrix elements and detunings follow from the definitions in Eqs. (12)-(14).

For the master-equation route, we evolve $\rho$ under the Lindblad equation in Eq. (15) with the dissipator definition in Eq. (16) and bath/linewidth relations in Eqs. (17)–(21) (and optional inhomogeneous broadening via a Voigt profile, discussed following the homogeneous-linewidth model). Steady states are obtained either by time evolution to convergence or by directly solving the steady-state condition $\dot\rho = 0$ together with $Tr(\rho) = 1$. The cavity Hilbert space is truncated at a photon cutoff N and increased until $n = \langle a^\dagger a \rangle$ and all reported observables are converged (relative change < $10^{-4}$). The two-emitter case uses the same procedure with a tensor-product TLS space and includes inter-dot exchange when needed (Eq. (39)).

For the rate-equation route (valid when $g \ll \kappa + \gamma_\perp$), we solve the reduced dynamics in Eqs. (23)-(24) using the exchange-balance relations in Eqs. (25)-(26), with the detuning-dependent exchange rate $\Gamma_c(\Delta)$ and filtered occupation defined in the main text (see the $\Gamma_c(\Delta)$ definition and the filtering discussed around the reduced model). We compute either the analytic near-resonance steady state (Eqs. (27)–(29)) and then assign effective temperatures via Eqs. (30)–(31), or the detuned steady state by solving the fixed-point conditions for $(\dot n, \dot p_e) = (0,0)$. For two emitters, we use the bright-channel description in the coherent collective limit (Eq. (47)); when mismatch and/or dephasing suppress emitter-emitter coherence, we switch to the independent-emitter system (Eqs. (48)-(50)).

We apply verification and validation tests throughout. All steady states satisfy $Tr(\rho) = 1$ and positivity within numerical tolerance. In the weak-coupling limit $g \to 0$ we recover the uncoupled thermal limits for the cavity and DQD baths, and in the strong-clamp limit the cavity reduces to the stream-like form discussed in the main text. At steady state we additionally verify exchange-balance consistency using Eq. (27), ensuring that energy currents between cavity, DQD, and baths balance. We confirm limiting behaviors (no baths, bath-pinned, emitter-pinned, and large-detuning suppression) and ensure numerical convergence by increasing N and tightening ODE tolerances until steady observables vary by < $10^{-4}$. Finally, we enforce the RWA/secular conditions stated in the text; when these conditions are not satisfied we rely on the full Lindblad model (and include additional terms as needed).

## 10 Design guidelines for experiments

This section summarizes practical parameter targets, quantitative expectations for $T_{\text{cav}}$ with representative numbers, and the main tradeoffs that set achievable performance in DQD-cavity refrigerators.

For operation near $\omega_c/2\pi \approx 5$ GHz, robust refrigeration is favored by large transverse coupling, weak bath loading of the cavity, and small homogeneous broadening of the DQD transition. Experimentally, the most direct lever for increasing the transverse exchange coupling is biasing near the DQD sweet spot ($\varepsilon \approx 0$), where the mixing angle approaches $\theta \approx \pi/2$ so that the transverse coupling is maximized ($g_\perp \approx g_0$) while the longitudinal component is suppressed. In what follows we set $g \equiv g_\perp$. A representative engineering target is $g/2\pi \sim 0.5 - 1.5\ MHz$.

Cooling performance is then controlled by the ratio of the exchange rate to the cavity loading rate, $\Gamma_c/\kappa$, so $\kappa$ should be minimized subject to measurement and stability constraints; values $\kappa/2\pi \lesssim$ 1-3 kHz are a realistic "strong-leverage" target for kelvin-stage operation when combined with aggressive filtering and careful port thermalization. The DQD homogeneous linewidth $\gamma_\perp$ should likewise be minimized (sweet-spot biasing reduces $\gamma_\phi$ and therefore $\gamma_\perp$); typical target values are $\frac{\gamma_\perp}{2\pi} \lesssim$ 0.2-0.5 MHz.

These scalings separate two regimes: if $\gamma_\perp \ll \kappa$ the exchange is cavity-limited, whereas if $\gamma_\perp \gg \kappa$ the exchange becomes dephasing-limited because $\Gamma_c \propto 1/(\kappa + \gamma_\perp) \approx 1/\gamma_\perp$.

Detuning should remain within the spectral-overlap window $|\Delta| \lesssim 2\Gamma$ so that $\Gamma_c(\Delta)$ and the filtered occupation entering Eq. (22) are not strongly suppressed. Finally, whether the device behaves stream-like or persistent is set by how the DQD is prepared: for stream-like operation one must impose a cold setpoint through a fast reset/clamp $\gamma_{\text{res}} \gg \Gamma_c, \kappa$, whereas for persistent operation one should avoid $\gamma_1$ so large that the DQD is pinned to $p_{\text{th}}$ and cannot act as a low-temperature reservoir.

Representative expectations at a 1 K stage can be stated directly in terms of thermal occupations. Here $\hbar\omega_c/k_B \approx 240$ mK, giving a bath occupation

$$\bar{n}_{\text{bath}}(1K) \approx \frac{1}{e^{0.24}-1} \approx 3.7. \tag{54}$$

If an engineered reset clamps the DQD to an effective setpoint $T_{\text{set}} \approx 50\text{-}80$ mK, then $n_{\text{set}} \approx 0.008\text{-}0.03$.

In the clamped (stream-like) limit, the steady cavity occupation is the weighted average

$$n^*(\Delta) = \frac{\kappa \bar{n}_{\text{bath}} + \Gamma_c(\Delta) n_{\text{set}}}{\kappa + \Gamma_c(\Delta)}, \qquad T_{\text{cav}} = \frac{\hbar\omega_c}{k_B \ln(1+1/n^*)}. \tag{55}$$

On resonance ($\Delta = 0$), representative cases illustrate the range of outcomes. With moderate exchange (e.g., $g/2\pi = 0.5$ MHz, $\kappa = 10$ kHz, $\gamma_\perp/2\pi = 0.3$ MHz), $\Gamma_c/\kappa$ can reach $\mathcal{O}(10^2\text{-}10^3)$, placing $T_{\text{cav}}$ typically in the $60-90$ mK range for $T_{\text{set}} \approx$ 50-80 mK. With stronger coupling and/or reduced $\kappa$ (e.g., $g/2\pi \approx 1$ MHz and $\kappa \approx 5$ kHz with $\gamma_\perp/2\pi \approx 0.2$ MHz), $\Gamma_c/\kappa$ readily exceeds $10^3$ and $T_{\text{cav}}$ approaches $T_{\text{set}}$ within a few mK. By contrast, in a dephasing-limited case (e.g., $\gamma_\perp/2\pi \approx 2$ MHz at otherwise similar $g$ and $\kappa$), $\Gamma_c(0)$ is reduced substantially, $\Gamma_c/\kappa$ can fall to $\mathcal{O}(10^1\text{-}10^2)$, and $T_{\text{cav}}$ typically rises into the $120-180$ mK range even with the same nominal $T_{\text{set}}$.

If both dots are active and symmetry is maintained, the bright-mode coupling is $g_B = |g_1 + g_2|$ (ideally $g_B = 2g$), so that $\Gamma_{c,B} \propto g_B^2 \approx 4\Gamma_c$, which can push single-mode refrigeration deeper into the sub-100 mK regime at fixed $\kappa$ and $T_{\text{set}}$. This gain is contingent on maintaining small mismatch and dephasing; once $|\Delta_{12}|$ or $\gamma_{\phi,j}$ becomes comparable to the effective linewidth, the enhancement collapses toward the independent-emitter sum (Sec. 6). At higher bath temperatures the basic requirement becomes more stringent because $\bar{n}_{\text{bath}}$ grows rapidly (e.g., $\bar{n}_{\text{bath}}(4K) \approx 8.6$, $\bar{n}_{\text{bath}}(10K) \approx 41$ at 5 GHz), so achieving $T_{\text{cav}} \ll T_{\text{bath}}$ demands both a very low DQD setpoint and very large $\Gamma_c/\kappa$; in practice, reducing the effective electromagnetic loading (infrared shielding, filtering, and careful thermalization) becomes as important as the intrinsic exchange parameters.

The dominant experimental tradeoffs follow directly from these scalings. Lowering $\kappa$ increases the leverage of exchange over bath loading, but it also lengthens ringdown times and tightens requirements on environmental shielding, microphonics, and the thermalization of all ports; a practical compromise is $\kappa/2\pi$ in the $1-3$ kHz range together with strong filtering and (when available) switchable coupling for calibration and readout. Collective (two-emitter) gain demands phase and frequency alignment: minimizing $\gamma_\phi$ through sweet-spot biasing and low-noise gating, and keeping $|\Delta_{12}| \ll \kappa + \bar{\gamma}_\perp$ so that a single bright resonance is preserved. If residual mismatch is unavoidable, a modest inter-dot exchange $\lambda$ can rehybridize the single-excitation manifold and partially brighten the nominally dark combination, but excessive $|\lambda|$ can split the resonance and reduce spectral overlap; optimal operation keeps the bright feature near $\Delta_B \approx 0$ without introducing splitting beyond the effective linewidth. Independently of the collective question, suppressing longitudinal coupling ($g_\parallel \ll g_\perp$) is important to avoid parasitic dispersive diffusion and excess dephasing; in practice this again points to operation near $\varepsilon \approx 0$ where $g_\parallel$ is minimized. Maintaining spectral overlap requires $|\Delta| \lesssim 2\Gamma$; slow drift can be compensated by servoing $\Delta$ via DQD spectroscopy or cavity-based indicators, while multimode operation can be managed by dedicating one "cold" mode to refrigeration and using controlled swaps to cool other modes as needed. For stream-like performance, the reset/clamp channel must dominate both cavity leakage and exchange; a practical target is

$$\gamma_{\rm res} \gtrsim 5 \times \max\{\kappa, \Gamma_c(\Delta)\} \tag{56}$$

Finally, $\Gamma_c(\Delta)$ and the relevant couplings can be calibrated in situ by comparing cavity ringdown with the DQD engaged/disengaged (effective $\kappa$ changes), by spectroscopy to extract linewidths and detuning dependence, and by dispersive shifts or photon-assisted tunneling to bound transverse and longitudinal coupling components.

A useful baseline to reach $T_{\rm cav} < 100$ mK at a $1$ K stage is $g/2\pi \gtrsim 0.7$-$1.0$ MHz, $\kappa/2\pi \lesssim 2$-$3$ kHz, $\gamma_\perp/2\pi \lesssim 0.2$-$0.4$ MHz, operation within $|\Delta| \lesssim \kappa + \gamma_\perp$ and $\theta \approx \pi/2$, together with an engineered reset that enforces $T_{\rm set} \lesssim 70$ mK. In this regime $\Gamma_c/\kappa$ typically reaches $\mathcal{O}(10^2$-$10^3)$, yielding $T_{\rm cav} \approx 60$-$90$ mK for a single active dot, and potentially lower when coherent two-emitter bright-mode enhancement is maintained. Overall, the design message is simple: maximize transverse coupling and spectral overlap, minimize $\kappa$ and $\gamma_\perp$, and provide a strong cold reset to impose a favorable setpoint; collective operation offers a quadratic gain in exchange rate but requires tight control of dephasing and frequency mismatch.

## 11 Results and Discussion

We now evaluate the cavity steady state predicted by the coarse-grained collision/persistent descriptions in terms of two directly measurable observables: the steady-state photon number $n^*$ and the corresponding effective cavity temperature $T_{\rm cav}$ defined as in Eq. (30). The results are organized to separate (i) spectral selectivity, controlled primarily by the detuning $\Delta$ between the relevant DQD transition frequency and the cavity frequency, from (ii) the relative weighting of engineered exchange versus phonon loading, controlled by $g$, the interaction time $\tau$ (via $\phi = g\tau$), the effective exchange rate $\Gamma_c$, and the cavity-phonon damping $\kappa$. This ordering is useful because $\Delta$ enters as a detuning filter in the effective exchange channel, while $\kappa$ sets the rate at which the cavity is pulled back toward the ambient phonon bath at $T_{\rm bath}$. We therefore begin with detuning sweeps, which isolate the role of resonance and linewidth in setting the steady state before introducing additional "knobs" such as $\kappa$ and $g$.

All results presented in this section follow from direct evaluation of the closed-form analytic expressions for the steady-state photon number derived in previous sections. To ensure consistency across parameter sweeps, a common baseline parameter set is adopted and summarized in Table 1. Unless otherwise stated in the figure captions or text (e.g., when sweeping a given control knob), all parameters are held at these baseline values.

Table 1: Global baseline parameters used in analytical evaluations

| Symbol | Baseline value | Meaning / where it enters |
|---|---|---|
| $T_{\rm bath}$ | 1.0 K | Thermal environment temperature loading the cavity through $\kappa$ |
| $T_{\rm set}$ | 50 mK | Reset/setpoint temperature defining the engineered DQD reservoir reference |
| $\omega_c$ | $2\pi \times 5 \times 10^9$ rad/s | Cavity angular frequency (5 $GHz$) |
| $\lambda$ | $2\pi \times 5 \times 10^9$ rad/s | Intra-pair exchange scale (collision-model reservoir internal Hamiltonian parameter) |
| $\tau$ | $50 \times 10^{-9}$ s | Interaction time per collision; sets small-angle exchange $\phi = g\tau$ |
| $R$ | $5 \times 10^6$ s$^{-1}$ | Poisson arrival rate (collision stream flux) |
| $\Delta$ | 0 rad/s | Baseline cavity-DQD detuning used when detuning is not being swept |
| $g$ | $2\pi \times 0.5 \times 10^6$ rad/s | Baseline transverse cavity-DQD coupling (0.5 MHz) |
| $\kappa$ | $2\pi \times 100$ rad/s | Cavity energy-damping rate to environment ("phonon tether") |
| $\chi$ | 2.0 | Two-dot collision enhancement factor via $\phi_{eff}^2 = \chi(g\tau)^2$ |
| $\gamma_1$ | $2\pi \times 1 \times 10^4$ rad/s | DQD longitudinal relaxation rate used in persistent-emitter modeling |
| $\gamma_\phi$ | $2\pi \times 0.3 \times 10^6$ rad/s | DQD pure dephasing rate; sets homogeneous linewidth $\gamma_\perp = \gamma_1/2 + \gamma_\phi$ |
| $\Delta_{12}$ | 0 rad/s | Inter-dot mismatch baseline used when mismatch is not swept |

Figure 1 shows $T_{\text{cav}}/T_{\text{dot}}$ and $n^*$ as functions of the DQD-cavity detuning $\Delta$ for the one-active-dot and two-active-dot interaction models. Panels (a,b) plot the temperature ratio, while panels (c,d) show the corresponding photon number.

In both models, the dominant trend is a resonance-centered modification of the cavity steady state. Near $\Delta = 0$ the DQD-mediated exchange channel is maximized, whereas for $|\Delta|$ larger than the relevant linewidth scale the exchange is spectrally suppressed and the cavity relaxes toward a bath-dominated steady state. This is the expected behavior of the effective overlap-controlled exchange rate: detuning reduces the ability of the DQD transition to absorb/emit cavity photons, so the engineered channel becomes ineffective and $\kappa$-weighted phonon loading increasingly determines $n^*$ and thus $T_{\text{cav}}$. Accordingly, Figure 1 (a,b) show the strongest departure from the far-detuned baseline near $\Delta = 0$, with a progressive loss of cooling as $|\Delta|$ increases; Figure 1 (c,d) exhibit the same crossover in $n^*(\Delta)$, with the smallest photon number near resonance and larger photon numbers as the cavity decouples from the engineered reservoir.

The qualitative distinction between the one- and two-active-dot cases is the sign and magnitude of the resonant shift in the effective detailed balance seen by the cavity. In the one-active-dot model (Figure 1a), the resonant steady state remains above the dot setpoint temperature, $\frac{T_{\text{cav}}}{T_{\text{dot}}} > 1$ across the sweep, with a representative minimum near ~4 for the plotted parameters. Thus resonance enhances exchange, but the net balance of upward and downward processes induced by the one-dot channel still corresponds to an effective hot setpoint relative to $T_{\text{dot}}$. In the two-active-dot model (Figure 1b), the resonant steady state crosses below unity, reaching $\frac{T_{\text{cav}}}{T_{\text{dot}}} < 1$ near $\Delta = 0$ for sufficiently large $g$ (down to ~0.8 in the plotted range). This sub-dot steady state is the signature of correlation-assisted refrigeration: the collective two-dot channel reshapes the effective excitation/relaxation balance (and therefore the effective setpoint) imposed on the cavity mode so that net photon extraction can dominate over bath loading in a finite detuning window.

The dependence on coupling strength g further clarifies the mechanism. Increasing g strengthens the weak interaction per collision ($\phi = g\tau \ll 1$) and thereby increases the DQD-induced Lindblad rates relative to $\kappa$, producing both (i) a deeper resonant minimum and (ii) a wider detuning bandwidth over which the engineered channel remains competitive. This trend is present in both models, but only the two-dot case converts increased coupling into genuine refrigeration ($\frac{T_{\text{cav}}}{T_{\text{dot}}} < 1$), whereas the one-dot case saturates above unity. The corresponding $n^*(\Delta)$ curves in Figure 1 (c,d) track these trends quantitatively: the resonant suppression of $n^*$ is stronger for larger $g$, and the recovery of $n^*$ at large $|\Delta|$ reflects the reversion to a bath-determined occupation as the exchange channel is detuning-filtered.

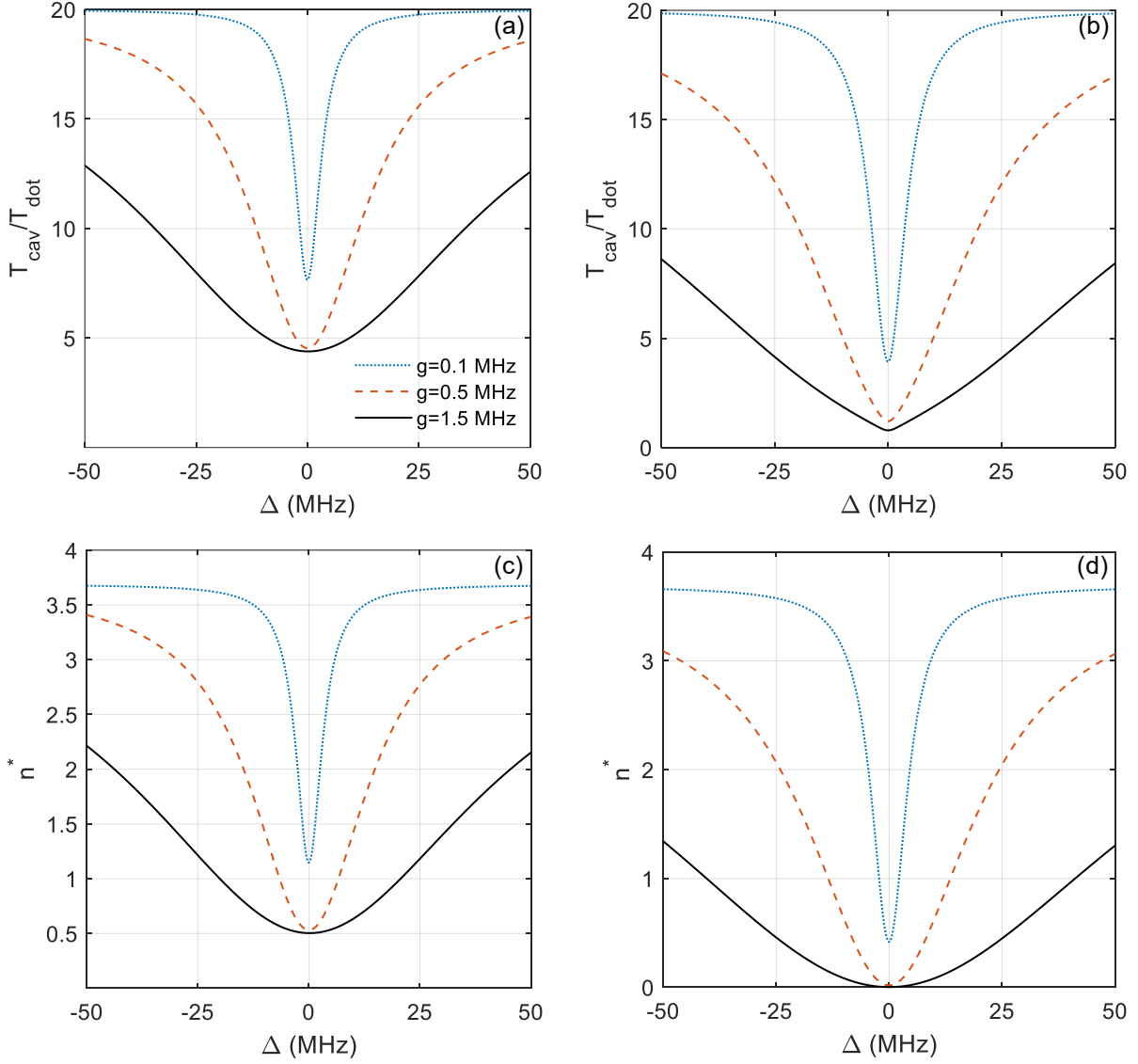

Figure 1: Detuning-controlled cooling and steady-state photon number are shown. (a,b) The effective cavity temperature, normalized to the dot (reset/setpoint) temperature, $T_{\text{cav}}/T_{\text{dot}}$, is shown versus the DQD-cavity detuning $\Delta = \omega_{qd} - \omega_c$ for the one-active-dot (a) and two-active-dot (b) interaction models. Cooling is strongest near resonance ($\Delta = 0$), where excitation exchange between the DQD transition and the cavity is maximized; at large $|\Delta|$ the exchange channel is spectrally suppressed, the phonon bath dominates, and the cavity relaxes toward a warmer steady state ($T_{\text{cav}} > T_{\text{dot}}$). Curves for three representative couplings $g$ illustrate the increased cooling depth and bandwidth with stronger coupling. In the one-dot model (a), the minimum temperature remains above the dot temperature ($\frac{T_{\text{cav}}}{T_{\text{dot}}} > 1$, with a minimum near $\sim 4$ for the parameters shown). In contrast, the two-dot model (b) exhibits correlation-assisted refrigeration, with $\frac{T_{\text{cav}}}{T_{\text{dot}}} < 1$ near resonance and at larger $g$ (down to $\sim 0.8$), demonstrating cavity cooling below the dot temperature. (c,d) The corresponding steady-state cavity occupation $n^*$ is plotted versus $\Delta$ for the one-dot (c) and two-dot (d) cases using the same parameters as in (a,b). The detuning dependence reflects spectral overlap: $n^*$ is reduced near resonance where energy exchange efficiently removes cavity excitations, and it increases for large $|\Delta|$ as the cavity decouples from the engineered reservoir and approaches the phonon-bath-limited occupation. The trends in $n^*(\Delta)$ directly mirror the temperature behavior in (a,b).

Figure 2 quantifies how the refrigeration performance is set by two independent "knobs" that enter the coarse-grained dynamics in distinct ways: (i) $\kappa$, which couples the cavity to an equilibrium bath at $T_{\text{bath}}$ and therefore fixes the strength of thermal repumping toward $\bar{n}_{\text{bath}}$, and (ii) $g$, which controls the engineered exchange rate through the weak-interaction parameter $\phi = g\tau$ (and thus the overall magnitude of the DQD-induced Lindblad terms). Interpreted at the level of the steady-state occupation, both knobs tune the relative weight of two competing setpoints: the bath-imposed setpoint (set by $\bar{n}_{\text{bath}}$) and the engineered-reservoir setpoint (set by the effective detailed-balance ratio associated with the DQD stream/refresh process).

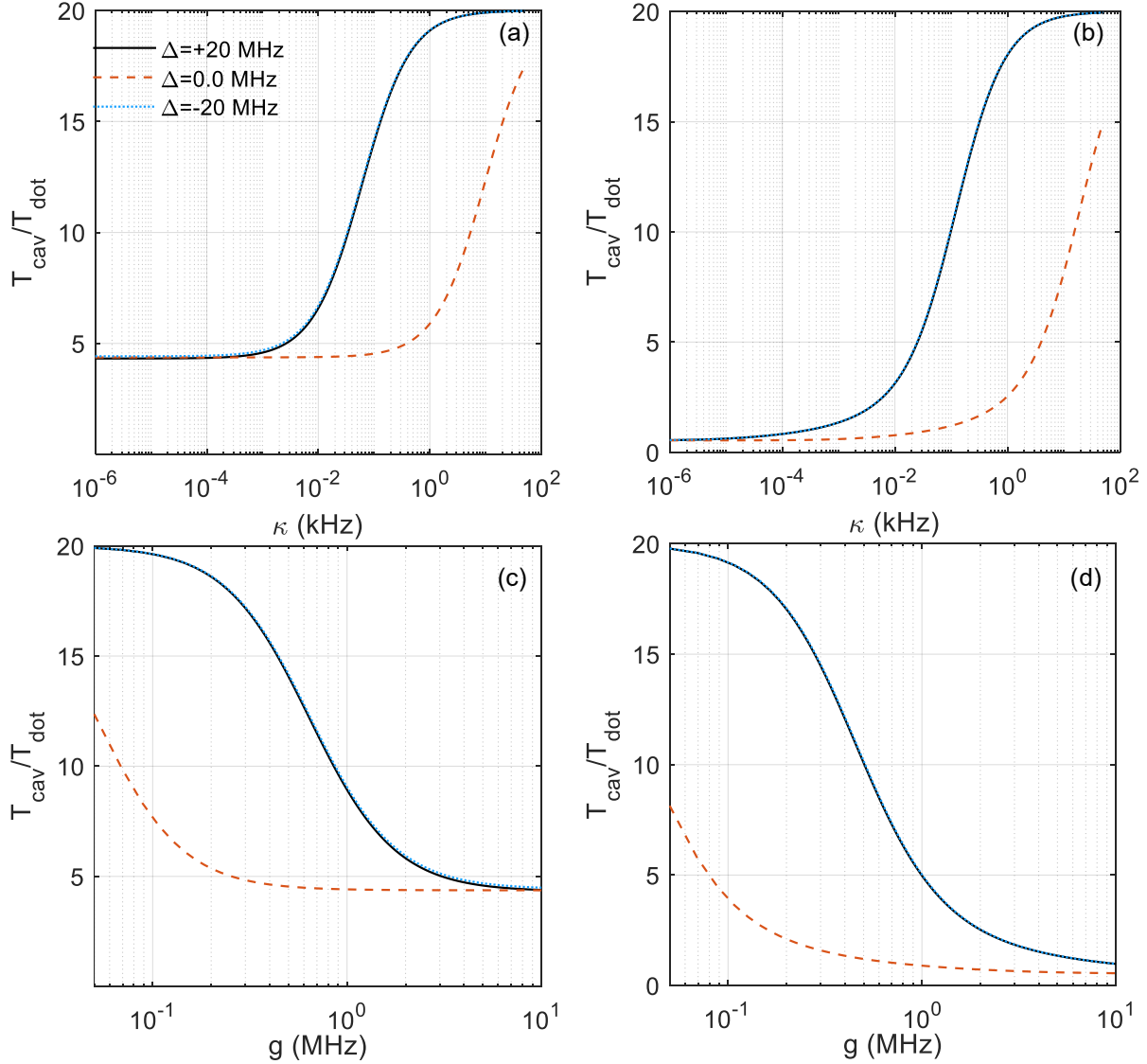

Figure 2: Refrigeration is shown versus the control knobs, phonon leakage and cavity-DQD coupling. (a,b) Steady-state cavity temperature, normalized to the dot (reset/setpoint) temperature, $T_{\text{cav}}/T_{\text{dot}}$, as a function of the cavity energy-damping rate to its environment (phonon tether) $\kappa$ for the one-active-dot (a) and two-active-dot (b) models. For weak leakage ($\kappa \to 0$), the engineered DQD reservoir sets the cavity steady state: the one-dot case saturates above the dot temperature ($\frac{T_{\text{cav}}}{T_{\text{dot}}} \approx 4$ for the parameters shown), whereas the two-dot case supports genuine refrigeration ($T_{\text{cav}}/T_{\text{dot}} \to 0.5$). As $\kappa$ increases, environmental damping progressively dominates the energy balance and drives the cavity back toward the thermal bath setpoint ($T_{\text{cav}} \to T_{\text{bath}}$), defining a crossover from reservoir-dominated to bath-dominated operation. (c,d) The same temperature ratio $T_{\text{cav}}/T_{\text{dot}}$ versus cavity-DQD coupling strength g for the one-dot (c) and two-dot (d) models. Increasing g strengthens the effective exchange per interaction ($\phi = g\tau \ll 1$), enhancing the engineered reservoir's ability to pull the cavity away from the bath-limited steady state and deepening the cooling. In both geometries the cooling ultimately saturates at the same limiting values as $\kappa \to 0$ ($\approx 4$ for one dot and $\approx 0.5$ for two dots), reflecting the intrinsic balance of upward and downward processes imposed by the engineered reservoir.

In Figure 2 (a) and (b) we vary $\kappa$ at fixed DQD parameters. In the $\kappa \to 0$ limit the bath contribution to both upward and downward cavity rates vanishes, so the steady state is determined predominantly by the engineered reservoir. This exposes the intrinsic "setpoint" imposed by each interaction geometry: the one-active-dot model saturates at $\frac{T_{\text{cav}}}{T_{\text{dot}}} \approx 4$ (Figure 2 a), whereas the two-active-dot model saturates at a much colder steady state, $T_{\text{cav}}/T_{\text{dot}} \to 0.5$ (Figure 2 b). This separation is not a small quantitative improvement; it reflects a qualitative change in the effective detailed balance seen by the cavity in the two-dot channel, which permits net photon extraction sufficient to cool the cavity below the reset temperature.

As $\kappa$ increases, the bath loading and the corresponding damping increasingly dominate the net rates, so the cavity is pulled back toward the equilibrium distribution at $T_{\text{bath}}$. The curves therefore exhibit a crossover from reservoir-dominated operation (small $\kappa$, where $T_{\text{cav}}$ is close to the engineered setpoint) to bath-dominated operation (large $\kappa$, where $T_{\text{cav}}$ approaches $T_{\text{bath}}$ and the ratio $T_{\text{cav}}/T_{\text{dot}}$ increases accordingly). Operationally, this crossover defines the leakage tolerance: it specifies the maximum $\kappa$ for which engineered cooling remains visible above the background thermalization channel.

Figure 2 (c) and (d) instead fix $\kappa$ and vary $g$. Increasing g increases $\phi = g\tau$ and therefore increases the engineered exchange contribution approximately as $\phi^2$ in the weak-collision regime. Consequently, at small g the engineered channel is too weak to compete with bath-induced loading and the cavity remains close to the bath-limited steady state. As $g$ is increased, the engineered rates grow and the cavity is progressively pulled away from the bath setpoint toward the reservoir-imposed setpoint, deepening the cooling and widening the parameter window where $T_{\text{cav}}/T_{\text{dot}}$ is minimized.

The saturation at large g in both panels is important: once the engineered channel dominates over $\kappa$, further increases in g no longer change the ratio of the engineered upward/downward processes, only their overall scale. The steady state then approaches the same limiting values revealed in Figure 2 (a,b) as $\kappa \to 0$: $\approx 4$ for the one-dot geometry and $\approx 0.5$ for the two-dot geometry. In this sense, $\kappa$ primarily controls how strongly the bath "anchors" the cavity, while g controls whether the engineered reservoir can overcome that anchoring; the asymptotic values, however, are set by the reservoir's internal detailed balance, which differs fundamentally between one- and two-dot operation.

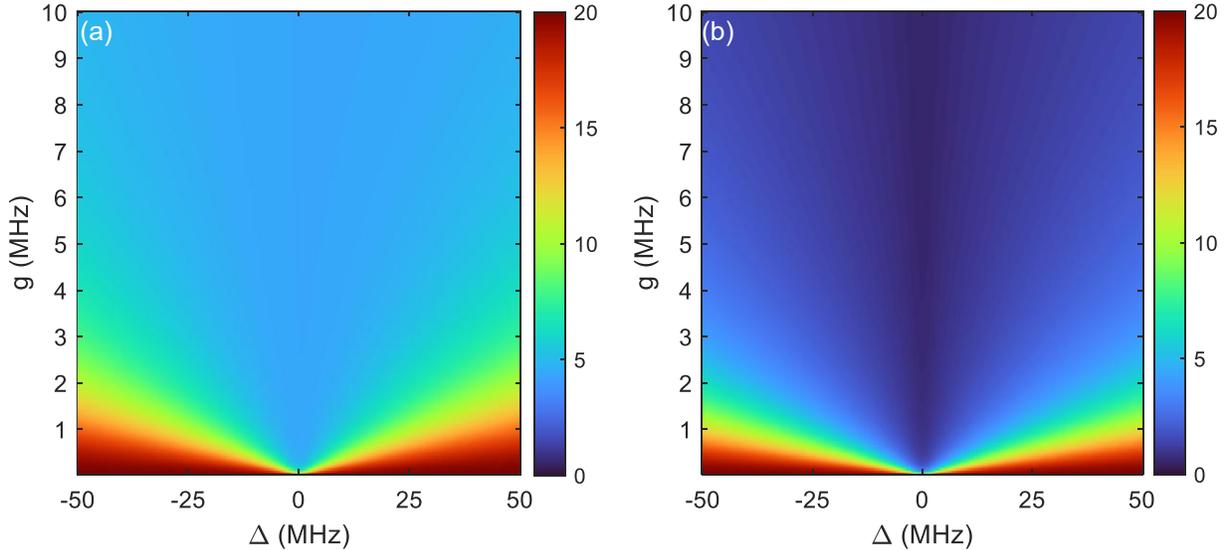

Figure 3: Two-dimensional cooling landscape versus detuning and coupling strength. (a,b) Color maps of the steady-state temperature ratio $T_{\text{cav}}/T_{\text{dot}}$ as a function of the DQD-cavity detuning $\Delta$ and transverse coupling strength $g$, shown for the one-active-dot model (a) and the two-active-dot model (b). In both cases, "cooling valleys" emerge near resonance ($\Delta \approx 0$) and deepen as $g$ increases, reflecting enhanced energy-exchange rates from stronger hybridization, while off-resonant ($|\Delta|$ large) or weak-coupling ($g$ small) regions revert toward bath-dominated behavior set by $\kappa$ and $T_{\text{bath}}$. The two-dot configuration exhibits substantially stronger refrigeration,

including an extended parameter region with $\frac{T_{\text{cav}}}{T_{\text{dot}}} < 1$, demonstrating cooling of the cavity mode below the dot (reset/setpoint) temperature.

Figure 3 maps the steady-state temperature ratio $T_{\text{cav}}/T_{\text{dot}}$ over the two primary control parameters of the collision-model dynamics, the DQD-cavity detuning Δ and the transverse coupling g, for the one-active-dot and two-active-dot geometries. In both panels of Figure 3, the lowest temperatures occur in a "valley" centered near resonance ($\Delta \approx 0$), where the spectral-overlap factor that controls the effective exchange channel is maximal and the DQD-induced upward/downward Lindblad rates are largest. Moving away from resonance suppresses the exchange channel, so the cavity steady state crosses over to bath-dominated behavior set by $\kappa$ and $T_{\text{bath}}$. Increasing g deepens and broadens the cooling valley because the engineered exchange scales with the weak-interaction parameter $\phi = g\tau$ (entering at leading order as $\phi^2$), thereby increasing the relative weight of the engineered reservoir compared to environmental damping.

The distinction between the two geometries is most clearly seen in the attainable minimum and the extent of the low-temperature region. For the one-active-dot model, the valley remains entirely above $\frac{T_{\text{cav}}}{T_{\text{dot}}} = 1$ across the mapped parameter range, consistent with the detuning cuts in Figure 1 and the limiting behavior in Figure 2. In contrast, the two-active-dot model exhibits a broad contiguous region with $\frac{T_{\text{cav}}}{T_{\text{dot}}} < 1$ near resonance and at moderate-to-strong coupling, indicating genuine refrigeration of the cavity below the reset/setpoint temperature. This two-dimensional view makes explicit that the two-dot advantage is not confined to a narrow fine-tuned point in parameter space; rather, it persists over an extended range of detunings and couplings where the engineered exchange channel dominates the bath-induced thermalization.

Figure 4 makes explicit how the "refreshing" assumption in the collision model can be related, in a controlled way, to a persistent-emitter implementation by using the dimensionless exchange-to-leakage ratio $\Gamma_c(0)/\kappa$ as a common x-axis. Here $\Gamma_c(0)$ denotes the on-resonance cavity-emitter exchange rate, set primarily by g and the homogeneous linewidth (via $\gamma_\perp$), while $\kappa$ captures energy leakage of the cavity into its environment. Increasing $\Gamma_c(0)/\kappa$ therefore moves the system from a bath-loaded cavity (small ratio) to a regime in which the engineered emitter channel dominates the energy balance (large ratio). Figure 4 (a) and Figure 4 (b) compare four steady-state descriptions at $T_{\text{bath}} = 1$ K: the refreshed-stream (collision) model (dashed), a persistent-emitter model with an imposed fast clamp/reset (solid), and a persistent-emitter model without clamp/reset for two intrinsic relaxation rates (dash-dotted $\gamma_1/2\pi = 1$ MHz, and dotted $\gamma_1/2\pi = 10$ kHz).

A key message is that "persistent" and "refreshed" reservoirs do not generically exhibit the same scaling with $\Gamma_c(0)/\kappa$, even when they share the same microscopic coupling to the cavity. In the no-clamp persistent model with fast intrinsic relaxation ($\gamma_1/2\pi = 1$ MHz), $T_{\text{cav}}$ closely tracks the clamped curve across the full range in both the one- and two-dot cases. This indicates that, for these parameters, the emitter effectively refreshes on its own: internal relaxation is rapid enough that correlations between successive exchange events are strongly suppressed, so the cavity experiences an approximately Markovian bath even without an explicit reset.

In contrast, when the emitter relaxes slowly ($\gamma_1/2\pi = 10$ kHz), the no-clamp curve separates visibly from the clamped limit and remains substantially warmer at large $\Gamma_c(0)/\kappa$ (e.g., $T_{\text{cav}} \approx 78\text{-}84$ mK when the clamped curve approaches $\approx 50$ mK). This behavior reflects memory and saturation in the coupled cavity-DQD dynamics: as exchange becomes fast compared to intrinsic emitter equilibration, the emitter no longer acts as an externally imposed setpoint reservoir for the cavity. Instead, the cavity and DQD partially thermalize with one another into a joint nonequilibrium steady state whose effective temperature is bounded away from the reset setpoint, so further increases in $\Gamma_c(0)$ do not yield proportional increases in net extraction of cavity excitations.

Introducing the clamp/reset (solid) restores the intended separation of timescales by actively re-preparing the emitter distribution between exchange events. As a result, $T_{\text{cav}}$ decreases monotonically with $\Gamma_c(0)/\kappa$ and approaches the reset-imposed setpoint at large ratio. In this clamped regime the one- and two-dot curves show broadly similar dependence on $\Gamma_c(0)/\kappa$ because the clamp primarily controls reservoir memory (the statistical independence of successive interactions), while the one- versus two-dot structure mainly sets the detailed-balance bias and the attainable temperature floor explored in Figure 1 to Figure 3.

Finally, the refreshed-stream (collision) prediction (dashed) exhibits the strongest scaling with $\Gamma_c(0)/\kappa$ because every interaction is with a newly prepared reservoir state, so the engineered rates remain strictly Markovian and continue to grow with effective interaction strength without degradation from memory. Within this idealized limit, the two-dot case shows a notably larger temperature drop than the one-dot case, consistent with collective enhancement and correlation-assisted modification of the effective upward/downward balance. At sufficiently large $\Gamma_c(0)/\kappa$, the two-dot stream curve can fall below the clamped persistent prediction, emphasizing that active reset is necessary to approach the refreshed limit and that a residual performance gap can persist at extreme exchange-to-leakage ratios when the persistent implementation cannot perfectly emulate instantaneous re-preparation.

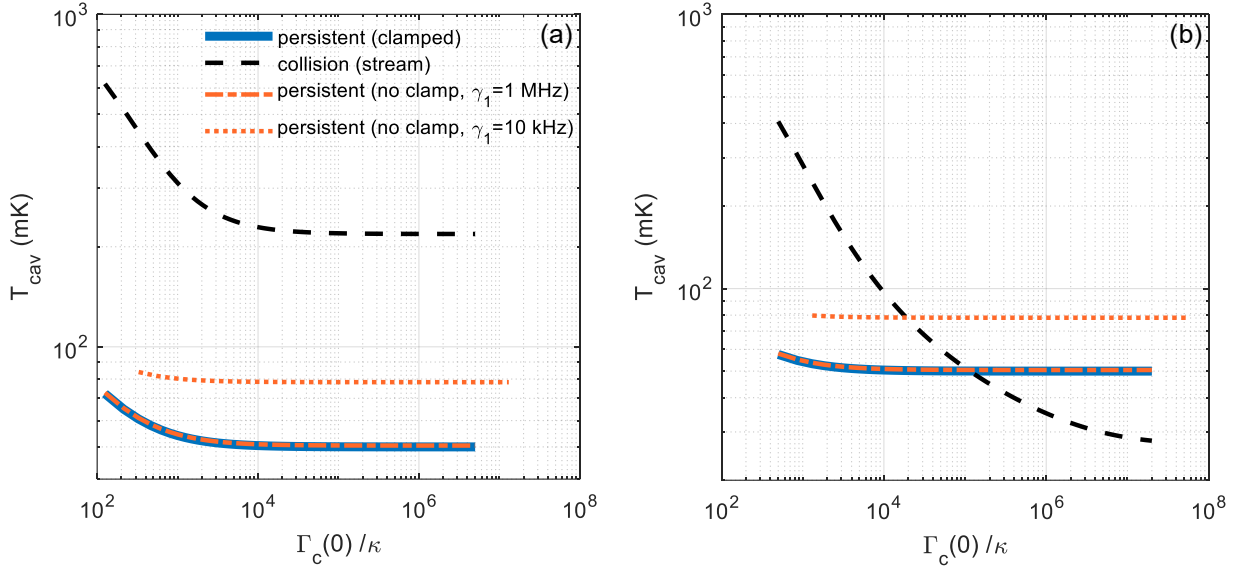

Figure 4: Mapping between persistent-emitter and refreshed-stream refrigeration via the exchange-to-leakage ratio is shown. (a,b) Steady-state cavity temperature $T_{\text{cav}}$ at $T_{\text{bath}} = 1$ K versus the on-resonance cavity-emitter exchange rate normalized by cavity leakage, $\Gamma_c(0)/\kappa$, for the one-active-dot (a) and two-active-dot (b) models. Curves compare a refreshed collision (stream) reservoir (dashed),[6] a persistent-emitter model with an explicit fast clamp/reset that enforces stream-like statistics (solid), and a persistent-emitter model without clamp/reset (dash-dotted and dotted lines). For the no-clamp model, two relaxation rates are shown to expose the role of emitter memory: with fast intrinsic relaxation ($\gamma_1/2\pi = 1$ MHz), $T_{\text{cav}}$ closely tracks the clamped result across the full range, indicating that the emitter effectively refreshes between exchanges; with slow relaxation ($\gamma_1/2\pi = 10$ kHz), $T_{\text{cav}}$ increases substantially and becomes only weakly improved by increasing $\Gamma_c(0)/\kappa$ (e.g., $T_{\text{cav}} \approx 78\text{-}84$ mK even when the clamped curve reaches $\approx 50$ mK), revealing saturation and memory in the coupled cavity-DQD dynamics. The clamped persistent model exhibits a monotonic reduction of $T_{\text{cav}}$ with increasing $\Gamma_c(0)/\kappa$ as the cavity is pulled from the bath-loaded steady state toward the reset-imposed setpoint $T_{\text{set}} = T_{\text{dot}} = 50$ mK, with similar trends for one and two active dots. The refreshed-stream prediction shows the strongest dependence on $\Gamma_c(0)/\kappa$ and the largest two-dot advantage; at sufficiently large $\Gamma_c(0)/\kappa$ the two-dot stream curve can fall below the clamped persistent curve, highlighting the performance gap between idealized refreshed reservoirs and practical persistent implementations at extreme exchange-to-leakage ratios.

Figure 5 quantifies how the refrigeration performance depends on the inter-dot frequency mismatch, $\Delta_{12} = \omega_{qd,1} - \omega_{qd,2}$, and, in the one-active-dot geometry, on the sign of that mismatch. The plotted observable is the steady-state cavity temperature $T_{\text{cav}}$ under otherwise fixed operating conditions, with the dashed curve corresponding to one active dot coupled to the cavity and the solid curve to two active dots coupled collectively.

In both geometries, increasing $|\Delta_{12}|$ suppresses cooling. Physically, mismatch reduces the spectral overlap needed for efficient cavity-DQD exchange and, in the two-dot setting, progressively destroys the collective (bright-mode) enhancement that underlies the strongest refrigeration. As $|\Delta_{12}|$ grows, the cavity dynamics revert toward a bath-dominated steady state because the engineered exchange channel becomes ineffective compared with $\kappa$.

The one-active-dot case is additionally sensitive to the sign of $\Delta_{12}$. Here, changing the sign does not simply "relabel" the dots, because only one dot is directly coupled to the cavity while the other dot enters indirectly through the internal two-dot level structure that sets the effective reservoir statistics. As a result, $T_{\text{cav}}(\Delta_{12})$ is not constrained to be even in $\Delta_{12}$. The observed asymmetry (lower $T_{\text{cav}}$ for $\Delta_{12} < 0$ than for $\Delta_{12} > 0$) reflects whether the cavity couples to the higher- or lower-energy dot relative to its partner: reversing the sign swaps which dot is energetically uphill within the pair, thereby changing the effective balance of upward versus downward processes that the cavity samples through the engineered reservoir.

By contrast, the two-active-dot model is symmetric under exchanging dots 1 and 2, so $T_{\text{cav}}(\Delta_{12})$ must be an even function of $\Delta_{12}$. The solid curve therefore depends only on $|\Delta_{12}|$ and captures the intrinsic fragility of the collective refrigeration mechanism to frequency nonuniformity, without any additional sign dependence.

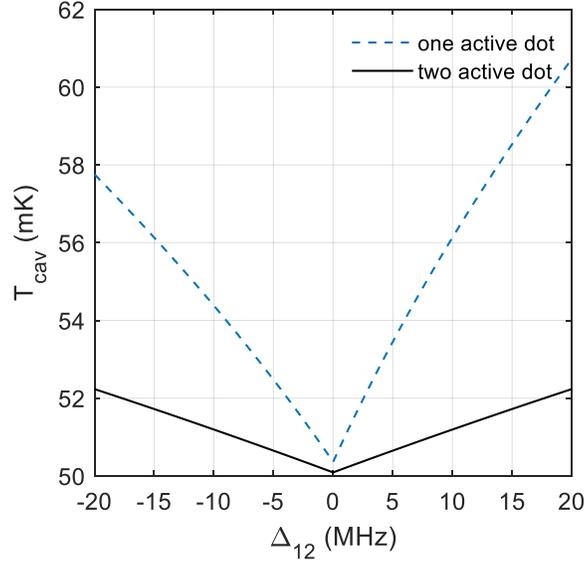

Figure 5: The sensitivity to dot-dot frequency mismatch and sign ($\Delta_{12}$) is shown. Steady-state cavity temperature $T_{\text{cav}}$ versus the signed inter-dot mismatch $\Delta_{12} = \omega_{qd,1} - \omega_{qd,2}$ for the one-active-dot model (dashed) and the two-active-dot model (solid). In both cases, increasing $|\Delta_{12}|$ degrades cooling because spectral alignment and collective exchange are progressively lost. The degradation is stronger in the one-active-dot case, where the refrigeration channel depends on a single dot's transition while the second dot primarily sets the internal level structure and effective reservoir statistics. In contrast, the two-active-dot model is symmetric under relabeling of the two dots, so $T_{\text{cav}}(\Delta_{12})$ is even in $\Delta_{12}$. The observed sign asymmetry in the one-active-dot curve ($T_{\text{cav}}$ lower for $\Delta_{12} < 0$ than for $\Delta_{12} > 0$) reflects whether the cavity couples to the higher- or lower-energy dot relative to its partner: changing the sign swaps which dot is energetically "uphill" in the pair, modifying the effective imbalance between excitation absorption and emission seen by the cavity.

Figure 6 isolates the role of bath loading by plotting the steady-state cavity temperature $T_{\text{cav}}$ versus the cavity energy-damping rate $\kappa$ for several bath temperatures $T_{\text{bath}}$, with the coupling fixed at $g/2\pi = 0.5$ MHz. This presentation makes the competition between the engineered DQD-induced exchange channel and the uncontrolled thermal leakage channel explicit: $\kappa$ simultaneously sets the rate at which energy is dumped into (and removed from) the cavity by the external environment and controls the bath-imposed occupation that the cavity would approach in the absence of engineered refrigeration.

In the $\kappa \to 0$ limit, the cavity is effectively decoupled from the external bath, so the steady state is set by the engineered reservoir alone. Both geometries therefore approach the same reset-imposed setpoint, $T_{\text{cav}} \approx T_{\text{set}} \approx 50$ mK, independent of $T_{\text{bath}}$, which confirms that the "cold" fixed point is not limited by the ambient bath but by the reservoir statistics encoded by the DQD/reset mechanism. As $\kappa$ increases, the bath increasingly loads the cavity with thermal photons; the steady state crosses over from reservoir-dominated to bath-dominated behavior and $T_{\text{cav}}$ rises toward $T_{\text{bath}}$. The ordering with $T_{\text{bath}}$ is therefore monotonic at fixed $\kappa$, and the separation between curves grows with $\kappa$ as the bath contribution becomes the dominant term in the energy balance.

Comparing Figure 6 (a) and (b), the rise of $T_{\text{cav}}$ with $\kappa$ is substantially steeper for the one-active-dot geometry. This reflects the weaker effective exchange channel in the one-dot case, so a smaller $\kappa$ is sufficient for bath loading to overwhelm the engineered refrigeration channel. In the two-active-dot configuration, collective enhancement increases the effective exchange rate and therefore shifts the crossover to larger $\kappa$, maintaining $T_{\text{cav}}$ near $T_{\text{set}}$ over a broader leakage range. Operationally, Figure 6 identifies the $\kappa$ window in which refrigeration remains meaningful for a given $T_{\text{bath}}$ and shows that the two-dot geometry enlarges this window, i.e., it is less sensitive to realistic bath coupling while preserving the same low-$\kappa$ setpoint.

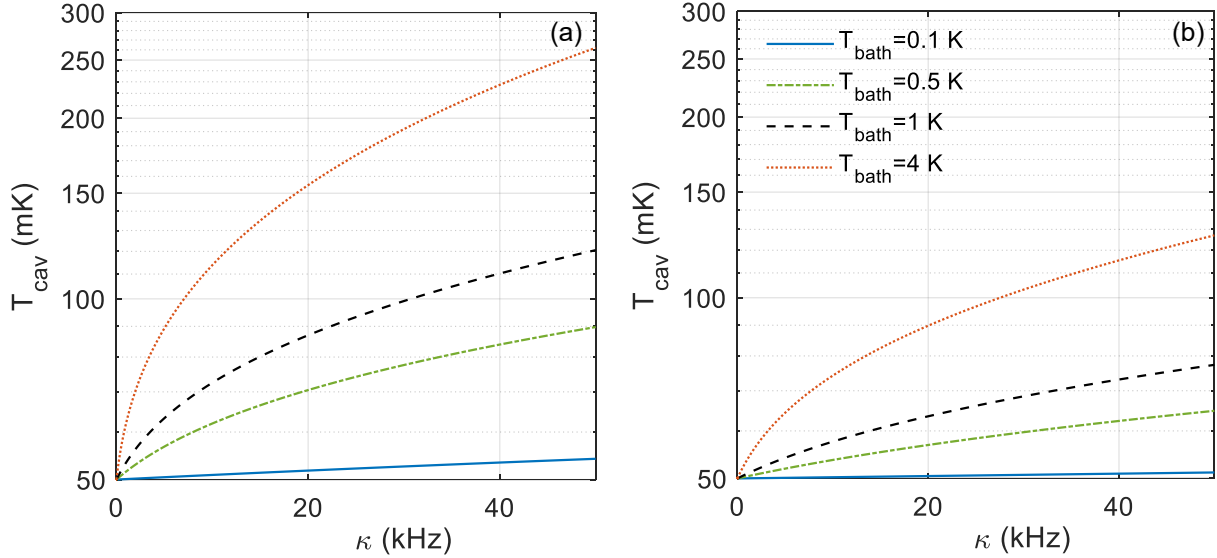

Figure 6: Bath-loading dependence of cavity cooling is shown. Steady-state cavity temperature $T_{\text{cav}}$ versus the cavity energy-damping rate $\kappa$ for several bath temperatures $T_{\text{bath}}$, shown for (a) one active dot and (b) two active dots at fixed coupling $g/2\pi = 0.5$ MHz. In the limit $\kappa \to 0$, the cavity is effectively isolated from the external bath and is driven to the engineered setpoint, $T_{\text{cav}} \approx T_{\text{set}} \approx 50$ mK, in both models. As $\kappa$ increases, thermal photons from the bath increasingly load the cavity and raise $T_{\text{cav}}$ toward $T_{\text{bath}}$. The rise is more rapid in the one-dot case because the effective exchange channel is weaker and less robust against bath loading, whereas the two-dot configuration provides an enhanced collective exchange that maintains cooling over a broader $\kappa$ range before the bath dominates.

## 12 Experimental implementation in cQED with DQD spin qubits

A practical bottleneck for scaling cQED hardware is increasingly the cryogenic heat budget rather than coherence in an isolated device. Wiring, attenuation, isolators/circulators, parametric amplifiers, and (increasingly) co-located control/readout electronics deposit heat and consume cooling power at the coldest stage, where the available cooling power is typically in the μW-$m$W range and drops rapidly with temperature. This is why many system-level roadmaps push functionality upward in the cryostat hierarchy ($1-4$ K stages) using cryogenic electronics and multiplexing, while reserving the base stage for only the most coherence-critical elements. Recent work on cryogenic CMOS and mixed-signal control for quantum systems explicitly targets this "4 K-class" integration regime, motivated by the heat-load and wiring constraints of large-scale machines.[10]

The difficulty, of course, is that microwave modes at a few GHz are thermally populated at kelvin temperatures: for a representative cavity frequency of $\omega_c/2\pi \approx 5$ GHz, the characteristic temperature is $\hbar\omega_c/k_B \approx 0.24$ K, giving a mean thermal occupation $\bar{n}_{\text{bath}} \approx 3.7$ at 1 K and $\bar{n}_{\text{bath}} \approx 16$ at 4 K. Thus, even if a superconducting cavity remains high-Q at these stages, the cavity field is "hot" in the sense relevant for dispersive shifts, measurement back-action, and photon-shot-noise dephasing. Experiments have long emphasized that even a small residual thermal photon population can measurably dephase superconducting circuits, motivating aggressive filtering/attenuation/thermalization strategies and careful accounting of nonequilibrium photon loading.[11,12]

This paper is motivated by the complementary, device-level question: instead of pushing the entire environment colder, can one locally refrigerate a specific microwave mode (or a small subset of modes) to sub-100 mK effective temperatures while the surrounding cryogenic stage remains at 1-4 K? The targeted objective here is not "ground-state cooling" in the optomechanics sense, but a controlled systematic reduction of the cavity's steady-state photon number and effective temperature, enough to suppress thermally driven errors, using an on-chip, actively resettable solid-state reservoir. In this context, gate-defined DQD spin qubits are appealing because (i) they naturally provide an electrically tunable spectrum and dipole coupling to microwave fields, and (ii) they support experimentally demonstrated circuit-QED style coupling to superconducting resonators across several material systems, forming a realistic pathway to an engineered, resettable reservoir embedded in the same microwave environment.[13]

## 12.1 Experimental mapping to a cQED architecture

Our analytic steady states are most usefully interpreted by mapping each model parameter onto a directly measurable (and, in many cases, in situ tunable) quantity in a circuit-QED device built around a superconducting microwave cavity coupled to a gate-defined DQD spin qubit.

Figure 7 summarizes the experimental architecture and the parameter mapping used throughout this work. The superconducting cavity (frequency $\omega_c$) is characterized by an effective energy-damping rate $\kappa$ that loads the mode from its environment at $T_{bath}$, while a gate-defined DQD provides a tunable solid-state reservoir. At the microscopic level, the DQD is an interacting two-electron system; in the half-filled, strong-interaction regime ($U \gg t_c$) the underlying Hubbard model reduces to an effective two-spin Hamiltonian (with exchange and SOC-induced anisotropic terms). The 'DQD transition frequency' $\omega_{qd,j}$ used throughout this paper refers to the relevant energy splitting of this reduced spin Hamiltonian for dot j (i.e., the effective two-level transition that couples transversely to the cavity), and therefore it can be renormalized by exchange/anisotropy in addition to the Zeeman scale. The key frequency coordinates are the cavity-DQD detunings $\Delta_j = \omega_{qd,j} - \omega_c$ and, in the two-emitter configuration, the inter-dot mismatch $\Delta_{12} = \omega_{qd,1} - \omega_{qd,2}$ that governs the preservation of collective bright/dark structure.

The cavity mode frequency $\omega_c$ is set by a 3D or planar resonator geometry (here we use a 5 GHz representative value consistent with many cQED implementations), while $\kappa$ is the net energy-damping rate that loads the cavity from its electromagnetic/phononic environment. In practice, $\kappa$ includes engineered coupling to a measurement line plus residual internal loss; it is therefore both device- and package-dependent, and it provides the primary "bath-loading knob" that interpolates between reservoir-dominated refrigeration ($\kappa$ small) and bath-dominated equilibration ($\kappa$ large). This separation of roles is precisely why $\kappa$ appears throughout the figures as the parameter controlling the crossover from a DQD-imposed nonequilibrium detailed balance to the ambient bath setpoint.

The DQD sector enters through (i) the cavity-DQD coupling strength $g$, (ii) the DQD homogeneous linewidth ($\gamma_1$ and $\gamma_\phi$, which set $\gamma_\perp = \gamma_1/2 + \gamma_\phi$), and (iii) the detunings. Experimentally, $g$ is not only a "device constant" but can be tuned over orders of magnitude by gate settings that change the DQD dipole moment (via charge admixture) and the spatial overlap to the cavity electric field. This is well established in semiconductor circuit QED experiments where gate-defined DQDs are integrated with superconducting resonators and operated from the dispersive to the strong-coupling regime.[14,15]

The detuning variable $\Delta$ is set by tuning the DQD level splitting $\omega_{qd}$ through gate detuning $\varepsilon$ and tunnel coupling t$_c$, which simultaneously controls the mixing angle (hence the relative longitudinal/transverse coupling weights) and the spectral overlap with the cavity. Operationally, $\Delta$ is a fast knob: in several DQD-cQED platforms, $\omega_{qd}$ can be shifted on nanosecond timescales by pulsing gate voltages or biasing along a detuning axis, enabling the "detuning-controlled cooling valleys" emphasized in Figure 1 and Figure 3 without modifying the cavity itself.[14,15]

For the two-emitter (two-active-dot) case, the additional control parameter is the inter-dot mismatch, $\Delta_{12}$. This mismatch is also gate-tunable, but in practice it is inevitably influenced by charge noise, device-to-device disorder, and slow drifts in the local electrostatic environment. The key architectural point is that $\Delta$ and $\Delta_{12}$ play distinct roles and remain largely independent knobs. The cavity detuning $\Delta$ primarily determines the spectral overlap of the cavity with the collective bright exchange channel, and therefore sets the strength and bandwidth of cavity-assisted energy exchange. By contrast, $\Delta_{12}$ controls how well the two emitters retain a collective bright/dark decomposition: when $\Delta_{12}$ is small, the symmetric (bright) combination couples efficiently to the cavity while the antisymmetric (dark) combination is protected; as $\Delta_{12}$ grows, the bright and dark sectors hybridize, the dark state is no longer dark, and the effective cooperativity advantage of the two-emitter configuration collapses. This separation of roles is why the two-dot benefit can appear "wide-area" in $(\Delta, g)$ maps (Figure 3), yet still show a sharp, threshold-like degradation as $\Delta_{12}$ increases (Figure 5).

In the device picture of Figure 7, the distinction between a refreshed-collision reservoir and a persistent emitter is implemented by the presence (or absence) of a fast reset/clamp pathway with rate $\Gamma_{\text{reset}}$. When $\Gamma_{\text{reset}}$ is large compared to the cavity-mediated exchange rate, each interaction event effectively samples an identically prepared DQD drawn from the same setpoint distribution at $T_{\text{set}}$, realizing the refreshed-collision reservoir assumed in the coarse-grained master-equation treatment. When $\Gamma_{\text{reset}}$ is absent or slow, the same DQD participates in many exchange events and its state evolves continuously under the competing influences of cavity exchange and its intrinsic rethermalization channels (phonons, charge noise, etc.), corresponding to the persistent-emitter regime. Framed this way, $\Gamma_{\text{reset}}$ is not merely a technical detail but a design knob that selects which thermodynamic picture applies and therefore which effective temperature and detailed-balance relations govern the coupled dynamics.

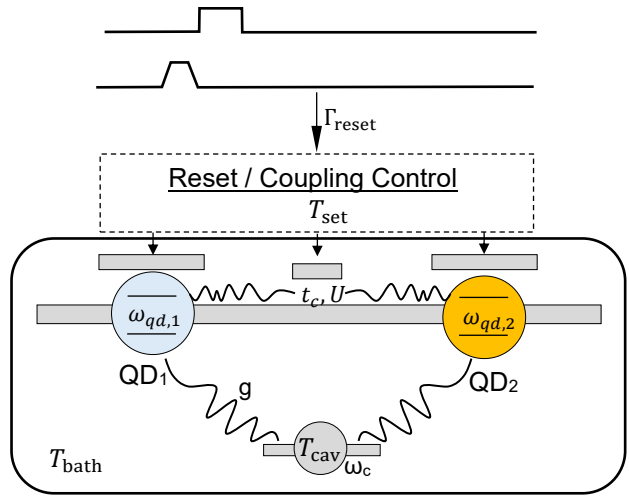

Figure 7: Experimental mapping of the refrigerator model to a cQED architecture using a gate-defined DQD spin qubit coupled to a superconducting microwave cavity. A single cavity mode at $\omega_c$ is loaded by an effective environmental damping $\kappa$ at an ambient stage temperature $T_{\text{bath}}$, while a gate-defined DQD provides an engineered reservoir whose transition frequencies $\omega_{qd,j}(j = 1,2)$ are tuned by gate detuning $\varepsilon$ and tunnel coupling $t_c$. The cavity-DQD spectral alignment is set by $\Delta_j = \omega_{qd,j} - \omega_c$, and the two-dot collective condition is controlled by the inter-dot mismatch $\Delta_{12} = \omega_{qd,1} - \omega_{qd,2}$. The one-active-dot configuration couples the cavity field primarily to a single dot ($g$), whereas the two-active-dot configuration couples to both dots, enabling bright/dark collective channels and correlation-assisted refrigeration. A fast reset/re-preparation pathway (rate $\Gamma_{\text{reset}}$) clamps the DQD to a reset/setpoint distribution at $T_{\text{set}}$, providing a controlled bridge between a persistent-emitter implementation and the refreshed-collision reservoir limit.

Achieving and maintaining a controlled reset temperature for selected quantum degrees of freedom in the millikelvin regime is experimentally realistic in modern quantum platforms, even when the surrounding environment is significantly warmer. A variety of active cooling strategies have been developed to selectively clamp specific subsystems, rather than the entire device, well below 100 mK. These include sympathetic cooling schemes demonstrated in trapped-ion systems,[16] sideband cooling protocols implemented in superconducting circuits and qubits,[17] and algorithmic cooling approaches that redistribute entropy within composite quantum systems.[18] While the physical implementations differ across platforms, these techniques collectively establish that targeted, subsystem-level cooling to millikelvin effective temperatures is achievable. In gate-defined double quantum dots specifically, microwave-induced cooling has been theoretically shown to suppress thermal fluctuations and stabilize spin populations at millikelvin temperatures relevant for qubit operation.[19] Together, these results provide a realistic experimental basis for the reset-setpoint temperature $T_{\text{set}} \approx 50\text{mK}$ assumed throughout this work and for operation in the fast-reset regime characterized by $\Gamma_{\text{reset}} \gg \Gamma_c, \kappa$.

Finally, even when the DQD itself is locally clamped to a milliKelvin reset setpoint, the broader cryogenic context motivating operation at elevated ambient temperatures ($\approx$1-4 K) is not simply convenience. It is driven by system-level heat budgets and wiring density, where the available cooling power and allowable dissipation at the coldest stage become the scaling bottleneck. Architectures that relocate control and support hardware to warmer stages (or co-integrate cryo-electronics near $1-4$ K) are therefore central to credible scaling roadmaps. This directly motivates a cavity-refrigeration primitive: an on-chip method to create local, sub-100 mK electromagnetic environments within a kelvin-stage setting, enabling qubit-adjacent photon and phonon engineering without requiring the entire platform to reside at the coldest temperature stage.[20]

## 12.2 DQD-engineered reservoir for the cavity

The refrigeration mechanism analyzed in Secs. 2-7 can be implemented in circuit QED by using a gate-defined DQD spin qubit as an actively resettable reservoir for a single cavity mode. The core idea is to replace the usual "static" bath seen by the cavity with a controllable, nonequilibrium detailed balance set by repeated, short exchange events between the cavity field and a DQD that is re-prepared to a setpoint state between events. This is the solid-state analogue of the micromaser/repeated-interaction construction used in the photo-Carnot model, where the reservoir is realized by a stream of identically prepared emitters.[5,6,21]

Each exchange event begins with a DQD prepared (by fast relaxation, tunneling to a reservoir, measurement-based reset, or gate-pulsed initialization) into a stationary state $\rho_{\text{DQD}}(T_{\text{set}})$ set by the effective spin Hamiltonian used throughout this work, $H_{\text{eff}} = \Delta_z (S_1^z + S_2^z) + J \mathbf{S}_1 \cdot \mathbf{S}_2 + \mathbf{D} \cdot (\mathbf{S}_1 \times \mathbf{S}_2) + \mathbf{S}_1 \cdot \overleftrightarrow{\Gamma} \mathbf{S}_2$, with parameters controlled by gates, magnetic field, and spin-orbit coupling. In practice, one does not require an ideal thermalizer; it is sufficient that the reset channel establishes a reproducible setpoint density matrix on a timescale $1/\Gamma_{\text{reset}}$ that is short compared with the interval between exchange events. The same DQD-cavity architecture and gate tunability needed here (control of $\omega_{qd}$, dipole coupling to the cavity field, and operation from dispersive to strong coupling) have been demonstrated across several semiconductor cQED platforms.[4,7,22,23]

During an exchange window of duration $\tau$, the relevant transition of the DQD that couples to the cavity mode (frequency $\omega_c$) can be treated as an effective two-level system with raising/lowering operators $\sigma_\pm$ defined in the dressed eigenbasis at the chosen working point. The interaction of each QD is then of Jaynes-Cummings form, $H_{\text{int}} = \hbar g(a\sigma_+ + a^\dagger \sigma_-)$, where $a$ is the cavity annihilation operator and $g$ is the tunable cavity-DQD coupling. In the repeated-interaction limit $\phi \equiv g\tau \ll 1$, expanding the unitary evolution to second order in $\phi$ and coarse-graining over a Poisson stream of exchange events with flux $R$ yields an effective Lindbladian for the cavity density operator $\rho_c$, $d\rho_c/dt = \kappa(\bar{n}_{\text{bath}} + 1)\mathcal{D}[a]\rho_c + \kappa\bar{n}_{\text{bath}}\mathcal{D}[a^\dagger]\rho_c + R\phi^2[r_1\mathcal{D}[a^\dagger]\rho_c + r_2\mathcal{D}[a]\rho_c]$. Here $\kappa$ is the cavity energy-damping rate to its ambient environment at $T_{\text{bath}}$, with $\bar{n}_{\text{bath}} = [\exp(\hbar\omega_c/k_B T_{\text{bath}}) - 1]^{-1}$, and $\mathcal{D}[L]\rho = L\rho L^\dagger - \{L^\dagger L, \rho\}/2$.

The reservoir enters only through the coefficients $r_1$ and $r_2$, defined as the excited- and ground-state populations of the specific DQD transition that couples to the cavity, evaluated in the prepared (reset/clamped) state: $r_1 = p_{A,e} = \text{Tr}[\Pi_{A,e}\,\rho_A], r_2 = p_{A,g} = \text{Tr}[\Pi_{A,g}\,\rho_A] = 1 - r_1$, where $\rho_A$ is the reduced density matrix of the active two-level degree of freedom and $\Pi_{A,e/g}$ are the corresponding projectors along its quantization axis. In the two-emitter configuration, an analogous identification applies to the collective bright mode in the symmetric limit; away from exact symmetry (mismatch/dephasing), we retain the two-emitter description rather than collapsing to a single pair $(r_1, r_2)$.

For $n = \langle a^\dagger a \rangle$, the above master equation implies $\dot{n} = -\kappa(n - \bar{n}_{\text{bath}}) + R\phi^2 [r_1(n+1) - r_2 n]$. The term proportional to $r_1$ describes photon creation (spontaneous plus stimulated emission, hence the factor $n+1$), while the term proportional to $r_2$ is photon annihilation (which requires an existing photon, hence the factor $n$). If the cavity were coupled only to the DQD stream ($\kappa \to 0$), the steady state ($\dot{n} = 0$) would be governed purely by the detailed-balance condition $r_1/r_2 = \exp(\hbar\omega_c/k_B T_{\text{set}})$, which defines the effective temperature $T_{\text{set}}$ imposed by the prepared DQD transition. With both the ambient bath (rate $\kappa$) and the DQD stream (characterized by $r_1$ and $r_2$) present, their competition determines the steady state. For $r_2 > r_1$ (the refrigeration regime), the closed-form solution for the cavity photon number is $n_{\text{ss}} = (\kappa\bar{n}_{\text{bath}} + R\phi^2 r_1)/(\kappa + $ $R\phi^2(r_2 - r_1)$). This expression makes the experimental requirements for active refrigeration transparent: (i) the DQD-induced exchange rate $R\phi^2(r_2 - r_1)$ must be comparable to or exceed $\kappa$; and (ii) the detailed-balance ratio $r_1/r_2$ must correspond to an effective temperature $T_{\text{set}} < T_{\text{bath}}$ at the chosen operating point (set by $\Delta$, $g$, and, in the two-active-dot case, $\Delta_{12}$).[6,22,23]

For practical implementation, the "refreshed-collision" regime assumed by the coarse-grained treatment is obtained when $\Gamma_{\text{reset}}$ is the dominant rate in the DQD sector (the DQD has no memory). This condition ensures each photon-exchange event begins with the DQD identically prepared in the state $\rho_{\text{DQD}}(T_{\text{set}})$. When $\Gamma_{\text{reset}}$ is reduced, the device crosses over to a persistent-emitter regime in which the DQD state co-evolves with the cavity field (the DQD retains memory across interactions with the cavity); this does not change the form of the cavity-DQD interaction, but it changes how $r_1$ and $r_2$ must be computed (from the joint steady state rather than from a fixed prepared $\rho_{\text{DQD}}$). In either case, the essential experimental knobs remain the same: $g$ and $\Delta$ via gate control of dipole admixture and level splitting, and $\Delta_{12}$ via relative tuning of the two dots in the two-emitter configuration.

### 12.3 Readout and validation metrics

The primary validation task is to extract (i) the steady-state cavity occupation $n = \langle a^\dagger a \rangle$ (and its associated effective temperature $T_{\text{cav}}$), and (ii) the effective temperature of the DQD degree of freedom participating in the exchange ($T_{\text{DQD}}$), under controlled sweeps of the detunings $\Delta_j$, coupling $g$, and (in the two-emitter configuration) mismatch $\Delta_{12}$. In practice, one does not need a unique definition of "temperature" to demonstrate refrigeration; it is sufficient to establish a reproducible reduction of $n$ relative to the equilibrium occupation set by $T_{\text{bath}}$, together with DQD population signatures consistent with energy flow from the DQD sector into the cavity and then out through $\kappa$.

The cavity occupation can be inferred directly from calibrated output noise spectroscopy: heterodyne detection of the cavity emission around $\omega_c$ yields a Lorentzian noise peak whose area is proportional to the intracavity photon number, after accounting for the known measurement chain gain and added noise. A complementary and often more sensitive approach uses an embedded qubit as a photon-number probe: residual thermal photons in a readout resonator produce measurable ac-Stark fluctuations and dephasing, allowing one to extract $n$ from time-domain qubit-based noise spectroscopy or dephasing-versus-power measurements. These methods have been used to distinguish coherent from thermal photon populations and to quantify extremely small residual occupations in circuit-QED cavities.[4,24,25] In the present context, "success" for cavity refrigeration is a steady $n$ that lies below the Bose-Einstein expectation $\bar{n}_{\text{bath}}(\omega_c, T_{\text{bath}})$, together with the predicted parameter dependence: $n$ should be minimized near the detuning-controlled cooling valleys and should degrade systematically as $\Delta_{12}$ spoils the collective bright/dark structure in the two-emitter case.

The relevant DQD metric is the population of the cavity-coupled transition (or the corresponding collective bright channel), which can be obtained from standard dispersive readout and/or reflectometry of the coupled DQD-resonator system. Gate-defined DQDs have demonstrated circuit-QED-style dispersive readout and microwave control, enabling extraction of excited-state probabilities $p_e$ via resonator phase/amplitude response, driven spectroscopy, or time-resolved relaxation measurements.[26,27]

An effective $T_{\text{DQD}}$ can then be assigned by mapping the measured populations onto the stationary distribution predicted by the DQD Hamiltonian at the working point (for a strict two-level reduction, $p_e = [1 + \exp(\hbar\omega_{qd}/k_B T_{\text{DQD}})]^{-1}$, for the full $H_{\text{eff}}$ model used here, $T_{\text{DQD}}$ is obtained by fitting the measured populations in the relevant eigenbasis to the Gibbs form or to the modeled steady state under reset). A strong consistency check is that, as $g$ is increased at fixed $\Delta$ (and in the refreshed-collision regime), $T_{\text{DQD}}$ should move toward $T_{\text{cav}}$ and the inferred energy flow should reverse sign when the cavity is deliberately heated above the DQD setpoint.

### 12.4 Operating regime checklist

The analysis relies on a separation of dynamical scales that can be summarized as the following experimental inequalities. For the refreshed-collision reservoir description, each exchange event must be weak and short, $\phi = g\tau \ll 1$, so that the coarse-grained Lindblad form with rates proportional to $R\phi^2$ is accurate and multiphoton exchange within a single event is negligible. To ensure that the DQD reservoir can pull the cavity away from equilibration to the ambient stage, the DQD-induced exchange channel must compete with the environmental loading of the cavity, i.e., the effective cavity pumping/relaxation scale set by the DQD stream (of order $R\phi^2(r_2 - r_1)$, times the spectral-overlap factor) should be comparable to or larger than $\kappa$. Finally, to cool the cavity-coupled DQD transition below $T_{\text{bath}}$ (and to pull $T_{\text{DQD}}$ toward $T_{\text{cav}}$), the cavity-mediated exchange rate should exceed the DQD's intrinsic rethermalization rate, $\Gamma_c(0) \gtrsim \gamma_1$, and operation should lie near the maximal-overlap window $|\Delta| \lesssim \Gamma$ (equivalently, within the dressed linewidth $2\Gamma$).

These conditions define the practical design space: $\kappa$ must be low enough and $g$ tunable enough that cavity-DQD exchange competes with bath loading, while reset/clamp pathways ($\Gamma_{\text{reset}}$) are fast enough to re-prepare the intended DQD state between exchange events.

**Conclusion**

We have presented a detuning-aware framework for using an exchange-coupled, gate-defined DQD spin-qubit system as an engineered reservoir for a superconducting microwave cavity that is otherwise loaded by a finite-temperature environment. Starting from an experimentally motivated effective two-spin Hamiltonian that captures SOC-induced anisotropic exchange, we constructed a practical mapping from device-level knobs (detuning, tunnel coupling, magnetic field orientation, and geometry) to the effective parameters that control cavity exchange: the cavity-DQD detuning, the exchange-enabled transverse matrix elements, and the single- and two-emitter statistics that set detailed balance. Incorporating ambient cavity loading through $\kappa$ and DQD homogeneous broadening through $\gamma_\perp$, we derived closed-form steady states solution for the cavity photon number and effective temperature, enabling rapid exploration of refrigeration "valleys" and parameter tradeoffs without relying on heavy numerical computation.

Two operating geometries were analyzed. In the one-active-dot configuration, where the cavity effectively couples to a single dot degree of freedom, the engineered channel can cool the cavity below $T_{\text{bath}}$ provided the cavity-DQD exchange rate competes with $\kappa$ and the DQD remains net absorptive ($p_g > p_e$) over the relevant transition. In the two-active-dot configuration, collective bright/dark (symmetric/antisymmetric) structure introduces an additional resource: correlated reservoir preparation can enhance the effective exchange channel and produce deeper refrigeration near resonance, while the emergence of a dark sector clarifies the conditions under which the enhancement collapses. We quantified the fragility of this collective advantage to inter-dot mismatch and dephasing, showing how increasing mismatch suppresses the effective bright-channel coherence and thereby reduces the overlap-limited exchange rate.

We also connected the refreshed-collision (stream) picture to a persistent-emitter implementation by identifying the exchange-to-loading ratio that controls when stream-like scaling is recovered and when memory effects lead to saturation. This bridge is useful experimentally: it translates the idealized reservoir assumptions of the collision model into concrete requirements on reset/clamp pathways $\Gamma_{\text{reset}}$ and on the hierarchy among exchange, decoherence, and bath loading. In this way, the theory supplies not only steady-state predictions but also an explicit operating-regime checklist that can be used to design and diagnose experiments.

From an implementation standpoint, the central message is that local electromagnetic refrigeration of a specific cavity mode is plausible even when the surrounding cryogenic stage is at elevated temperature ($\approx 1-4$ K), provided that $\kappa$ is sufficiently small and that $g$ and the detuning can be tuned into the overlap window where $\Gamma_c$ dominates. The mapping and validation metrics discussed in the experimental section suggest a direct route to testing the mechanism: measure the cavity thermal occupation via standard cQED thermometry (e.g., sideband/noise thermometry or qubit-assisted photon counting) while independently characterizing DQD linewidths and reset dynamics, and compare the extracted $n^*$ and $T_{\text{cav}}$ against the analytic predictions across $(\Delta, g, \kappa)$ and mismatch sweeps.

Looking forward, several extensions are immediate. First, a fully microscopic treatment of the DQD reset and rethermalization channels (phonons, charge noise, and driven reset protocols) would enable quantitative predictions for $\Gamma_{\text{reset}}$ and for the effective reservoir temperature under realistic duty cycles. Second, integrating the present cavity refrigeration primitive with a downstream use case, such as creating locally cold microwave environments for qubit readout or suppressing photon-shot-noise dephasing, would sharpen the system-level impact and clarify the most relevant performance metrics. Finally, the same formalism can be generalized to multi-mode and multi-reservoir networks, where engineered frequency conversion and reservoir preparation could create spatially localized cold spots and entropy flow pathways in complex cryogenic architectures.

In summary, by combining an experimentally grounded DQD Hamiltonian, a detuning-resolved overlap description, and analytic steady states, we provide a compact and design-oriented theory for cavity refrigeration using DQD spin-qubit reservoirs. The resulting parameter maps, mismatch sensitivity, and operating inequalities should help guide near-term cQED experiments aimed at realizing locally sub-100 mK microwave modes within kelvin-stage cryogenic environments.

**Acknowledgement**

This study is partially based on work supported by AFOSR under contract number FA9550261B011, and by the NSF under grant number CBET-2110603.

**APPENDIX: Mapping rules from microscopic DQD parameters to effective model quantities**

This appendix provides a constructive mapping between microscopic DQD parameters and the effective quantities appearing in the minimal S/T Hamiltonian. Explicit expressions are given for the local effective fields $\mathbf{B}_{1,2}(\mathbf{B}, \mathbf{D}, \overleftrightarrow{\Gamma})$ and for the Dzyaloshinskii-Moriya (DM) axis rotation angle $\alpha$, which at leading order satisfies $\tan\alpha \approx |\mathbf{D}|/J$.

**A.1 DM axis rotation (Shekhtman transformation)**

We define the unit vector along the DM interaction as

$$\hat{\mathbf{n}} = \frac{\mathbf{D}}{|\mathbf{D}|}$$

Opposite rotations are applied to the two spins according to $\mathbf{S}'_1 = R(\hat{\mathbf{n}}, +\alpha)\,\mathbf{S}_1$, $\mathbf{S}'_2 = R(\hat{\mathbf{n}}, -\alpha)\,\mathbf{S}_2$, where $R(\hat{\mathbf{n}}, \pm\alpha) = \exp\left[-i(\pm\alpha)\hat{\mathbf{n}} \cdot \mathbf{S_1}\right]$ for spin 1 and $\exp[-i\,(\pm\alpha)\,\hat{\mathbf{n}} \cdot \mathbf{S}_2]$ for spin 2 (equivalently, $R$ acts on the corresponding spin subspace). Here $\hat{\mathbf{n}}$ is taken along the DM vector, and the angle $\alpha$ is chosen such that $\cos\alpha = \frac{J}{J^0}$, $\sin\alpha = |\mathbf{D}|/J_0$, $J_0 \equiv \sqrt{J^2 + |\mathbf{D}|^2}$.

With this choice, the combined Heisenberg and DM interaction transforms as

$$J\,\mathbb{I} + [\mathbf{D}]_\times \;\to\; J_0\,\mathbb{I},$$

where $[\mathbf{D}]_\times$ is the antisymmetric $3 \times 3$ matrix with components

$$\left([\mathbf{D}]_\times\right)_{ij} = \varepsilon_{ijk}\,D_k$$

where $\varepsilon_{ijk}$ is the Levi-Civita symbol (with $\varepsilon_{123} = +1$)

After rotation, the Hamiltonian takes the form

$$H' = J_0\,\mathbf{S}'_1 \cdot \mathbf{S}'_2 + \mathbf{S}'_1 \cdot \overleftrightarrow{\Gamma'} \cdot \mathbf{S}'_2 + g^*\mu_B(\mathbf{B}_1 \cdot \mathbf{S}'_1 + \mathbf{B}_2 \cdot \mathbf{S}'_2),$$

with the transformed symmetric anisotropy tensor

$$\overleftrightarrow{\Gamma'} = R(\hat{\mathbf{n}}, +\alpha)\,\overleftrightarrow{\Gamma}\,R(\hat{\mathbf{n}}, -\alpha)^T.$$

If the original Hamiltonian contains only Heisenberg and DM terms $\left(\overleftrightarrow{\Gamma} = 0\right)$, the transformation generates a residual symmetric anisotropy of order $\mathcal{O}(|\mathbf{D}|^2/J)$, which may be absorbed into $J_0$ or treated explicitly as

$$\overleftrightarrow{\Gamma}_{\mathrm{DM}} \simeq (J_0 - J)\,\hat{\mathbf{n}} \otimes \hat{\mathbf{n}}$$

The rotation matrix appearing above is given by Rodrigues' formula,

$$R(\hat{\mathbf{n}}, \alpha) = \cos\alpha\,\mathbb{I} + \sin\alpha\,[\hat{\mathbf{n}}]_\times + (1 - \cos\alpha)\,\hat{\mathbf{n}}\hat{\mathbf{n}}^T$$

**A.2 Explicit local fields $\mathbf{B}_1, \mathbf{B}_2$ from a uniform laboratory field**

The Zeeman term transforms according to

$$g^*\mu_B \mathbf{B} \cdot (\mathbf{S}_1 + \mathbf{S}_2) \;\to\; g^*\mu_B(\mathbf{B}_1 \cdot \mathbf{S}'_1 + \mathbf{B}_2 \cdot \mathbf{S}'_2),$$

where the local effective fields are

$$\mathbf{B}_1 = R(\hat{\mathbf{n}}, +\alpha)\, \mathbf{B},\; \mathbf{B}_2 = R(\hat{\mathbf{n}}, -\alpha)\, \mathbf{B}$$

Using Rodrigues' formula, these can be written explicitly as

$$\begin{aligned}\mathbf{B}_1 &= \cos\alpha\, \mathbf{B} + \sin\alpha\, (\hat{\mathbf{n}} \times \mathbf{B}) + (1-\cos\alpha)(\hat{\mathbf{n}} \cdot \mathbf{B})\, \hat{\mathbf{n}}, \\ \mathbf{B}_2 &= \cos\alpha\, \mathbf{B} - \sin\alpha\, (\hat{\mathbf{n}} \times \mathbf{B}) + (1-\cos\alpha)(\hat{\mathbf{n}} \cdot \mathbf{B})\, \hat{\mathbf{n}}\end{aligned}$$

It is useful to introduce the average and half-difference fields,

$$\begin{aligned}\bar{\mathbf{B}} &= \frac{\mathbf{B}_1 + \mathbf{B}_2}{2} = \cos\alpha\, \mathbf{B} + (1-\cos\alpha)(\hat{\mathbf{n}} \cdot \mathbf{B})\, \hat{\mathbf{n}}, \\ \Delta\mathbf{B} &= \frac{\mathbf{B}_1 - \mathbf{B}_2}{2} = \sin\alpha(\hat{\mathbf{n}} \times \mathbf{B})\end{aligned}$$

The inhomogeneous component therefore originates from the part of $\mathbf{B}$ transverse to $\hat{\mathbf{n}}$. Writing

$$\mathbf{B}_\perp = \mathbf{B} - (\hat{\mathbf{n}} \cdot \mathbf{B})\, \hat{\mathbf{n}},$$

its magnitude satisfies

$$|\Delta\mathbf{B}| = |\mathbf{B}_\perp|\, \sin\alpha$$

### A.3 Minimal S/T parameters $\delta$ and $k$ from $\mathbf{B}_1, \mathbf{B}_2$

Choosing the S/T quantization axis along the laboratory $z$ direction, projection of the Zeeman term onto the basis $\{|T_+\rangle, |T_0\rangle, |T_-\rangle, |S_0\rangle\}$ yields

$$\begin{aligned}\delta &= \frac{g^*\mu_B}{2}(B_{1,z} + B_{2,z}) = g^*\mu_B\, (\bar{\mathbf{B}} \cdot \hat{\mathbf{z}}), \\ k &= \frac{g^*\mu_B}{\sqrt{2}}(\Delta\mathbf{B} \cdot \hat{\mathbf{y}}_{\text{eff}}),\end{aligned}$$

where $\hat{\mathbf{y}}_{\text{eff}}$ denotes the transverse direction that mixes $|S_0\rangle$ with $|T_\pm\rangle$.

Substituting the explicit fields gives:

$$\begin{aligned}\delta &= g^*\mu_B\, [\cos\alpha\, B_z + (1-\cos\alpha)(\hat{\mathbf{n}} \cdot \mathbf{B})\, \hat{n}_z], \\ k &= \frac{g^*\mu_B}{\sqrt{2}}\, \sin\alpha\, (\hat{\mathbf{n}} \times \mathbf{B}) \cdot \hat{\mathbf{y}}_{\text{eff}}\end{aligned}$$

where $\hat{n}_z$ is the $z$-component of $\hat{\mathbf{n}}$. A convenient choice is to align $\hat{\mathbf{y}}_{\text{eff}}$ with the $y$ component of $\hat{\mathbf{n}} \times \mathbf{B}$, leading to

$$k = \frac{g^*\mu_B}{\sqrt{2}}\, \sin\alpha\, |(\hat{\mathbf{n}} \times \mathbf{B})_y|$$

In the special configuration $\hat{\mathbf{n}} \perp \mathbf{B}$ with $\hat{\mathbf{z}} \parallel \mathbf{B}$, one finds

$$\delta = g^*\mu_B B \cos\alpha, \qquad k = \frac{g^*\mu_B}{\sqrt{2}} B \sin\alpha,$$

which reproduces the compact parametrization used in Eq. (7) of the main text with $\beta_y = \sin\alpha$.

### A.4 Unitary map between the S/T basis and the product basis

In several places in the main text we (i) construct the thermal (or reset-setpoint) two-spin state in the eigenbasis of the minimal S/T Hamiltonian and then (ii) evaluate single-dot populations by partial tracing in the product basis. For completeness, we state the basis ordering and the explicit unitary connecting these representations.

We order the S/T basis as:

$$|\chi_{ST}\rangle = (|T_+\rangle, |T_0\rangle, |T_-\rangle, |S_0\rangle)^T,$$

with

$$|T_+\rangle = |\uparrow\uparrow\rangle, \quad |T_-\rangle = |\downarrow\downarrow\rangle, \quad |T_0\rangle = \frac{|\uparrow\downarrow\rangle + |\downarrow\uparrow\rangle}{\sqrt{2}}, \quad |S_0\rangle = \frac{|\uparrow\downarrow\rangle - |\downarrow\uparrow\rangle}{\sqrt{2}}$$

We order the product basis as

$$|\chi_{\text{prod}}\rangle = (|\uparrow\uparrow\rangle, |\uparrow\downarrow\rangle, |\downarrow\uparrow\rangle, |\downarrow\downarrow\rangle)^T$$

The unitary that maps S/T coordinates to product coordinates (columns are the S/T basis vectors expressed in the product basis) is:

$$U \equiv U_{\text{prod}\leftarrow\text{ST}} = \begin{pmatrix} 1 & 0 & 0 & 0 \\ 0 & 1/\sqrt{2} & 0 & 1/\sqrt{2} \\ 0 & 1/\sqrt{2} & 0 & -1/\sqrt{2} \\ 0 & 0 & 1 & 0 \end{pmatrix}.$$

Thus, density matrices in the two representations are related by

$$\rho_{\text{prod}} = U \rho_{\text{ST}} U^\dagger, \qquad \rho_{\text{ST}} = U^\dagger \rho_{\text{prod}} U$$

This is the transformation used before tracing out the inactive dot and evaluating the "active-dot" populations that define $r_1$ and $r_2$.

The thermal density matrix of the DQD is given by:

$$\rho_{\text{ST}} = \begin{pmatrix} \rho_{11} & 0 & \rho_{13} & \rho_{14} \\ 0 & \rho_{22} & 0 & 0 \\ \rho_{13}^* & 0 & \rho_{33} & \rho_{34} \\ \rho_{14}^* & 0 & \rho_{34}^* & \rho_{44} \end{pmatrix},$$

where,

$$\rho_{11} = \frac{1}{Z}\sum_{j=0}^{2} e^{-\beta\lambda_{j+2}}(N_{j+2})^{-2}|A_{j+2}|^2,$$

$$\rho_{22} = \frac{1}{Z}e^{-\beta J_0/4},$$

$$\rho_{33} = \frac{1}{Z}\sum_{j=0}^{2} e^{-\beta\lambda_{j+2}}(N_{j+2})^{-2}|B_{j+2}|^2,$$

$$\rho_{44} = \frac{1}{Z}\sum_{j=0}^{2} e^{-\beta\lambda_{j+2}}(N_{j+2})^{-2},$$

$$\rho_{13} = \frac{1}{Z}\sum_{j=0}^{2} e^{-\beta\lambda_{j+2}}(N_{j+2})^{-2} A_{j+2} B_{j+2}^*,$$

$$\rho_{14} = \frac{1}{Z}\sum_{j=0}^{2} e^{-\beta\lambda_{j+2}}(N_{j+2})^{-2} A_{j+2},$$

$$\rho_{34} = \frac{1}{Z}\sum_{j=0}^{2} e^{-\beta\lambda_{j+2}}(N_{j+2})^{-2} B_{j+2},$$

and the partition function $Z$ is:

$Z = e^{-\beta J_0/4} + \sum_{j=0}^{2} e^{-\beta\lambda_{j+2}}$. Here, $\beta \equiv 1/(k_B T_{\text{bath}})$, and the parameters $A_j$, $B_j$, $N_j$, and $\lambda_j$ are defined in the subsection 3.1. The density matrix of the DQD in the product basis is given by:

$$\rho_{\text{prod}} = \begin{pmatrix} \rho_{11} & \rho_{14}/\sqrt{2} & -\rho_{14}/\sqrt{2} & \rho_{13} \\ \rho_{14}^*/\sqrt{2} & \frac{\rho_{22}+\rho_{44}}{2} & \frac{\rho_{22}-\rho_{44}}{2} & \rho_{34}^*/\sqrt{2} \\ -\rho_{14}^*/\sqrt{2} & \frac{\rho_{22}-\rho_{44}}{2} & \frac{\rho_{22}+\rho_{44}}{2} & -\rho_{34}^*/\sqrt{2} \\ \rho_{13}^* & \rho_{34}/\sqrt{2} & -\rho_{34}/\sqrt{2} & \rho_{33} \end{pmatrix}.$$

In the product basis, the matrix elements satisfy $\langle\uparrow\uparrow|\rho_{\text{prod}}|\uparrow\downarrow\rangle = -\langle\uparrow\uparrow|\rho_{\text{prod}}|\downarrow\uparrow\rangle$, indicating antisymmetric correlations between the two QDs. Due to this sign difference, the two reduced density matrices, $\rho_A$ and $\rho_B$ are distinct:

$$\rho_A = \text{Tr}_B[\rho_{\text{prod}}] = \begin{pmatrix} \rho_{11} + \frac{\rho_{22}+\rho_{44}}{2} & (\rho_{34}^* - \rho_{14})/\sqrt{2} \\ (\rho_{34} - \rho_{14}^*)/\sqrt{2} & \rho_{33} + \frac{\rho_{22}+\rho_{44}}{2} \end{pmatrix},$$

$$\rho_B = \text{Tr}_A[\rho_{\text{prod}}] = \begin{pmatrix} \rho_{11} + \frac{\rho_{22}+\rho_{44}}{2} & -(\rho_{34}^* - \rho_{14})/\sqrt{2} \\ -(\rho_{34} - \rho_{14}^*)/\sqrt{2} & \rho_{33} + \frac{\rho_{22}+\rho_{44}}{2} \end{pmatrix}.$$

While the off-diagonal elements differ in sign, the diagonal elements of the reduced density matrices are identical. Consequently, this distinction does not affect the single-emitter statistics when $\hat{\mathbf{w}}_1$ is aligned with the product-basis quantization axis (so only populations enter). For general $\hat{\mathbf{w}}_1$, use Eq. (10), which depends on the full reduced state. We define these populations as: $p_e = r_1 = \rho_{33} + \frac{\rho_{22}+\rho_{44}}{2}$ (population of the excited state $|\downarrow\rangle$), and $p_g = r_2 = \rho_{11} + \frac{\rho_{22}+\rho_{44}}{2}$ (population of the ground state $|\uparrow\rangle$).

In the two-active-dot geometry, the two-emitter statistics are given in terms of the matrix elements of $\rho_{\text{prod}}$ as: $r_1^{(2)} = \rho_{33} + \rho_{22}$, $r_2^{(2)} = \rho_{11} + \rho_{22}$.

### A.5 Local quantization axes for the active dot

The local axes defining the single-dot populations are

$$\hat{\mathbf{w}}_1 = \frac{\mathbf{B}_1}{|\mathbf{B}_1|}, \qquad \hat{\mathbf{w}}_2 = \frac{\mathbf{B}_2}{|\mathbf{B}_2|}$$

These axes enter the single-emitter probabilities through

$$p_{A,e} = \frac{1}{2} + \frac{1}{2} \text{Tr}\big[(\boldsymbol{\sigma}_1 \cdot \hat{\mathbf{w}}_1)\rho_{\text{prod}}\big], \qquad p_{A,g} = 1 - p_{A,e},$$

after thermalization of the pair and partial tracing over the inactive dot.

### A.6 Spectral-overlap exchange rate and filtered cavity occupation (derivation of Eqs. (23)-(24))

This subsection derives the detuning-filtered cavity-DQD energy exchange rate and the resulting closed-form expression for the cavity steady-state photon number used in the main text.

### A.6.1 Starting master equation and weak-exchange reduction

Consider one cavity mode $a$ (frequency $\omega_c$) coupled to a single effective DQD transition with lowering operator $\sigma_-$ and transition frequency $\omega_{qd}$. In an interaction picture with respect to the uncoupled Hamiltonian, the coherent exchange is

$$H_{\text{int}} = \hbar g\,(a\sigma_+ + a^\dagger \sigma_-),$$

with detuning $\Delta \equiv \omega_{qd} - \omega_c$. The open-system dynamics include (i) cavity loading from the ambient environment at $T_{\text{bath}}$ with energy damping rate $\kappa_{\text{eff}}$ and thermal occupation

$$\bar{n}_{\text{bath}} = \left[\exp\left(\frac{\hbar\omega_c}{k_B T_{\text{bath}}}\right) - 1\right]^{-1},$$

and (ii) homogeneous broadening of the DQD transition captured by longitudinal relaxation $\gamma_1$ and pure dephasing $\gamma_\phi$, so that the transverse decoherence rate is

$$\gamma_\perp = \frac{\gamma_1}{2} + \gamma_\phi$$

The full Lindblad equation is

$$\frac{d\rho}{dt} = -\frac{i}{\hbar}[H_{\text{int}}, \rho] + \kappa(\bar{n}_{\text{bath}} + 1)\,\mathcal{D}[a]\rho + \kappa\bar{n}_{\text{bath}}\,\mathcal{D}[a^\dagger]\rho + \gamma_1\,\mathcal{D}[\sigma_-]\rho + \frac{\gamma_\phi}{2}\,\mathcal{D}[\sigma_z]\rho.$$

To isolate the cavity dynamics in the reservoir-engineering regime, we adopt the prepared-reservoir approximation used in the main text: during each exchange window the populations of the cavity-coupled DQD transition are clamped to their prepared values, $p_e = p_{A,e} = \text{Tr}[\Pi_{A,e}\rho_A]$, $p_g = p_{A,g} = \text{Tr}[\Pi_{A,g}\rho_A] = 1 - p_{A,e}$, so that the corresponding inversion is fixed by the reset/setpoint preparation. Under weak exchange ($g$ small enough that the exchange coherence relaxes rapidly compared with the cavity photon-number dynamics), one can adiabatically eliminate the exchange coherence $\langle a^\dagger \sigma_- \rangle$ to obtain an effective rate description.

### A.6.2 Elimination of the exchange coherence and Lorentzian overlap

Define the cavity photon number $n \equiv \langle a^\dagger a \rangle$ and the exchange coherence

$$C \equiv \langle a^\dagger \sigma_- \rangle$$

Using the master equation and standard operator identities for Lindblad generators, one finds the coupled equations (keeping only terms up to second order in $g$ in the final reduced dynamics):

$$\frac{dn}{dt} = -\kappa(n - \bar{n}_{\text{bath}}) + ig(C - C^*),$$
$$\frac{dC}{dt} = -\left[\frac{\kappa + \gamma_\perp}{2} + i\Delta\right] C + ig[p_e(n+1) - p_g n]$$

In the adiabatic-elimination limit for the coherence ($dC/dt \approx 0$), we obtain

$$C \approx \frac{ig\,[p_e(n+1) - p_g n]}{(\kappa + \gamma_\perp)/2 + i\Delta}$$

Substituting into $dn/dt$ gives

$$\frac{dn}{dt} = -\kappa(n - \bar{n}_{\text{bath}}) + 2g^2 \,\text{Re}\left\{\frac{1}{(\kappa+\gamma_\perp)/2 + i\Delta}\right\} [p_e(n+1) - p_g n].$$

The real part evaluates to a Lorentzian spectral-overlap factor,

$$\text{Re}\left\{\frac{1}{\frac{\kappa+\gamma_\perp}{2} + i\Delta}\right\} = \frac{\frac{\kappa+\gamma_\perp}{2}}{\Delta^2 + [(\kappa+\gamma_\perp)/2]^2}$$

It is therefore convenient to define the detuning-filtered exchange rate

$$\Gamma_c(\Delta) \equiv \frac{4g^2(\kappa+\gamma_\perp)}{(\kappa+\gamma_\perp)^2 + 4\Delta^2}$$

With this definition, the photon-number equation becomes

$$\frac{dn}{dt} = -\kappa(n - \bar{n}_{\text{bath}}) + \Gamma_c(\Delta)\,[p_e(n+1) - p_g n]$$

This is the explicit "spectral-overlap" result used in the main text: the cavity-DQD exchange enters only through the Lorentzian overlap between the cavity linewidth $\kappa_{\text{eff}}$ and the DQD homogeneous linewidth $\gamma_\perp$.

### A.6.3 Filtered steady-state photon number and effective cavity temperature

Rearranging the above equation yields a linear rate form

$$\frac{dn}{dt} = -\Gamma_\downarrow n + J_\uparrow,$$

with

$$\Gamma_\downarrow = \kappa + \Gamma_c(\Delta)\,(p_g - p_e), \qquad J_\uparrow = \kappa \bar{n}_{\text{bath}} + \Gamma_c(\Delta)\,p_e$$

Hence the steady-state photon number is:

$$n^* = n_{ss} = \frac{\kappa \tilde{n}_{\text{bath}} + \Gamma_c(\Delta)\, p_e}{\kappa + \Gamma_c(\Delta)(p_g - p_e)}$$

This reproduces the closed-form steady-state used for the "clamped" (prepared-reservoir) cavity temperature in the main text. The associated effective cavity temperature $T_{\text{cav}}$ is obtained from the Bose inversion

$$T_{\text{cav}} = \frac{\hbar \omega_c}{k_B} \frac{1}{\ln\left(1 + \frac{1}{n^*}\right)}$$

The physical conditions for refrigeration are transparent in this form: one requires (i) $p_g > p_e$ so that the DQD induces net damping rather than gain, and (ii) $\Gamma_c(\Delta)$ comparable to or larger than $\kappa$ so that the engineered exchange channel competes with ambient loading.

### A.7 Two-emitter bright/dark reduction and collective exchange rate (derivation of Eq. (47))

This subsection derives the two-active-dot collective exchange rate used in the main text and clarifies how inter-dot mismatch $\Delta_{12}$ and homogeneous broadening reduce the enhancement.

#### A.7.1 Bright/dark operators and enhanced coupling

For two active dots ($j = 1,2$) coupled to the same cavity mode, the exchange interaction is

$$H_{\text{int}} = \hbar g [a(\sigma_+^{(1)} + \sigma_+^{(2)}) + a^\dagger(\sigma_-^{(1)} + \sigma_-^{(2)})]$$

Define symmetric (bright) and antisymmetric (dark) operators

$$S_+^B = \frac{\sigma_+^{(1)} + \sigma_+^{(2)}}{\sqrt{2}}, \quad S_+^D = \frac{\sigma_+^{(1)} - \sigma_+^{(2)}}{\sqrt{2}}, \quad S_-^{B,D} = (S_+^{B,D})^\dagger,$$

$g_B \equiv \sqrt{2} g$, so:

$$H_{\text{int}} = \hbar g_B (a S_+^B + a^\dagger S_-^B),$$

#### A.7.2 Where the additional factor-of-2 comes from (cross correlators)

The relevant static factor contains

$$\langle (\sigma_1^- + \sigma_2^-)(\sigma_1^+ + \sigma_2^+) \rangle = \langle \sigma_1^- \sigma_1^+ \rangle + \langle \sigma_2^- \sigma_2^+ \rangle + \langle \sigma_1^- \sigma_2^+ \rangle + \langle \sigma_2^- \sigma_1^+ \rangle.$$

#### A.7.3 Incorporating mismatch as bright/dark mixing and effective broadening

The mismatch and average detuning are

$$\Delta_{12} \equiv \omega_{qd,1} - \omega_{qd,2}, \qquad \Delta \equiv \frac{\omega_{qd,1} + \omega_{qd,2}}{2} - \omega_c$$

Define the exchange coherences

$$C_B \equiv \langle a^\dagger S_-^B \rangle, \qquad C_D \equiv \langle a^\dagger S_-^D \rangle$$

Their equations of motion are

$$dC_B/dt = -A\, C_B - i(\Delta_{12}/2) C_D + i g_B F_B,$$
$$dC_D/dt = -A\, C_D - i(\Delta_{12}/2) C_B + i g_B F_D,$$

where $A \equiv \frac{\kappa + \gamma_\perp}{2} + i\Delta$ in the symmetric limit.

Eliminating $C_D$ at steady state gives, in general,

$$C_B = [i g_B F_B A + (\Delta_{12}/2) g_B F_D]/[A^2 + (\Delta_{12}/2)^2].$$

If the dark channel is not directly driven ($F_D \approx 0$), this reduces to

$$C_B \approx (i g_B F_B)/(A + \frac{(\Delta_{12}/2)^2}{A})$$

At $\Delta = 0$ this denominator is $\bar{\Gamma} + (\Delta_{12}/2)^2/\bar{\Gamma}$ with $\bar{\Gamma} = (\kappa + \bar{\gamma}_\perp)/2$, i.e., mismatch suppresses the bright response similarly to an increased effective linewidth.)

### A.7.4 Collective bright-channel exchange rate used in the main text (Eq. 47)

For two active dots ($j = 1, 2$), define $\Delta_j = \omega_{qd,j} - \omega_c$ and the bright-channel detuning $\Delta_B \approx (\Delta_1 + \Delta_2)/2$. In the symmetric limit $g_1 = g_2 = g$ and matched transition frequencies, the cavity couples only to the bright combination $S_-^B = (\sigma_1^- + \sigma_2^-)/\sqrt{2}$, giving an effective single-emitter interaction $H_{int} = \hbar\, g_B\, (a\, S_+^B + a^\dagger S_-^B)$ with $g_B = \sqrt{2} g$ (see A.7.1). Adiabatic elimination of the exchange coherence proceeds exactly as in the single-emitter case (Appendix A.6), with $\gamma_\perp$ replaced by an effective bright-channel transverse rate $\bar{\gamma}_\perp$, yielding the Lorentzian-overlap exchange rate quoted in the main text [Eq. (47)].

Frequency mismatch $\Delta_{12}$ and additional dephasing reduce collectivity by mixing bright and dark coherences (A.7.3). At the rate-equation level this is captured by treating $\bar{\gamma}_\perp$ as an effective broadening (mismatch/noise dependent), which suppresses $\Gamma_{c,B}$ through the same overlap denominator in Eq. (47).

### A.8 Practical mapping summary (from device parameters to the effective model)

This section summarizes the chain of mappings used in the main text; the detailed formulas are given in the cited equations/subsections.

1. Device controls → effective DQD transition and mixing (Secs. 3.2, Eqs. (12)-(14))
From ($\varepsilon, t_c$) obtain $\omega_{qd}$ and the mixing angle $\theta$; this sets the transverse/longitudinal cavity couplings ($g_\perp, g_\parallel$) and the detuning $\Delta = \omega_{qd} - \omega_c$.

2. Microscopic spin physics → rotated (DM-free) two-spin Hamiltonian (Appendix A.2-A.3; main-text Eqs. (6)-(8))
From exchange and SOC-induced anisotropy obtain $J_0$ and the DM rotation parameters ($\hat{n}, \alpha$). These determine the rotated local fields $B_1, B_2$ and hence the local dot axes $\hat{w}_1, \hat{w}_2$.

3. Two-spin thermal state → effective "reservoir statistics" (Appendix A.4; main-text Eqs. (10), (34))
Thermalize the two-spin Hamiltonian at $T_{bath}$ to get $\rho_{ST}$, transform to the product basis, and trace out as needed. The single-emitter statistics entering the cooling theory are $r_1 = p_{A,e} = Tr[\Pi_{A,e}\, \rho_A]$ and $r_2 = 1 - r_1$ (with $\Pi_{A,e}$ defined in Eq. (10)). The two-emitter case uses the corresponding collective bright/dark basis (Appendix A.7; main-text Eqs. (45)-(50)).

4. Exchange rate used in reduced dynamics (Sec. 4; Eqs. (3), (47))
Given g (or $g_B$ in the bright limit) and the relevant transverse linewidth (γ_⊥ or γ̄_⊥), the cavity-emitter exchange rate is $\Gamma_c(\Delta)$ [Eq. (3)] for one emitter and $\Gamma_{c,B}(\Delta_B)$ [Eq. (47)] for the collective bright channel.